\documentclass[aps,prc,noshowpacs,showkeys,nofootinbib]{revtex4}

\usepackage{color} 

\usepackage{graphicx}
\usepackage{dcolumn}
\usepackage{amsthm}
\usepackage{bm}

\usepackage{isotope}
\usepackage{braket}
\usepackage{physics}

\newcommand{\sigmabf}{\mbox{\boldmath $\sigma$}}

\begin{document}

\title{Bremsstrahlung emission from nuclear reactions in compact stars
}

\author{Sergei~P.~Maydanyuk$^{(1)}$}%
\email{maidan@kinr.kiev.ua}%

\author{Kostiantyn~A.~Shaulskyi$^{(2)}$}%
\email{konstiger1998@live.com}%

\affiliation{$(1)$Institute for Nuclear Research, National Academy of Sciences of Ukraine, Kyiv, 03680, Ukraine}
\affiliation{$(2)$Taras Shevchenko National University of Kyiv, Ukraine}

\date{\small\today}



\begin{abstract}
Bremsstrahlung emission of photons during nuclear reactions inside dense stellar medium is investigated in the paper.
For that, a new model of nucleus is developed, where nuclear forces combine nucleons as bound system in dependence on deep location inside compact star.
A polytropic model of stars at index $n=3$ with densities characterized from white dwarf to neutron star is used.
Bremsstrahlung formalism and calculations are well tested on existed experimental information for scattering of protons of light nuclei in Earth.
%
%
We find the following.
(1) In neutron stars a phenomenon of dissociation of nucleus is observed --- its disintegration on individual nucleons, starting from some critical distance between this nucleus and center of star with high density.
%
%
We do not observe such a phenomenon in white dwarfs.
%
(2) In the white dwarfs, influence of stellar medium imperceptibly affects on bremsstrahlung photons.
Also, we have accurate description of bremsstrahlung photons in nuclear reactions in Sun.
(3) For neutron stars, influence of stellar medium is essentially more intensive and it crucially changes the bremsstrahlung spectrum.
The most intensive emission is from bowel of the star, while the weakest emission is from periphery.
%
%
%
\end{abstract}



\maketitle

\section{Introduction
\label{sec.introduction}}

Light from stars provides us a main information about them.
Here, photons emitted due to processes with participation of elementary particles in stellar medium have been studied the most intensively.
But, we know that stars have nuclei from the lightest up to heavy (for example, see Chapter IV in book~\cite{Lewis.2004.book} for nuclei in Sun).
Stability of calculations of the bremsstrahlung spectra is achieved, if to take into account space regions up to atomic shells and larger, even for low energies.
This indicates on importance to study nucleus as unite system of evolving nucleons, which should be studied via solution of many body quantum mechanical problem, where interactions cannot be small and study cannot be reduced to perturbation approaches.
So, important ingredient is emission of photons from stars produced in nuclear reactions with participation of nuclei, from the lightest up to heavy.
Bremsstrahlung emission of photons during nuclear reactions in dense stellar medium has not been studied deeply and, so, it is a topic of current paper.

Study of nuclear forces~\cite{Epelbaum.2009.RMP,Tilley.2002.NPA} in extreme conditions enlarges our possibilities to understand them deeper, where stars is a good laboratory for investigations~\cite{Bisnovatyi-Kogan.2011.book,Bisnovatyi-Kogan.1989.book,Kippenhahn.2012.book}.
Quantum nature of nuclear interactions can be displayed in possibility of nucleons to combine as bound system, i.e. atomic nucleus, characterized via binding energy.
%
Many-nucleon unified theories of nucleus and nuclear reactions
(for example, see microscopic cluster models based on developments of
resonating group method~\cite{Wildermuth.1977.book,Tang.1978.PR,Tang.1981.lectures}
and generator coordinate method~\cite{Horiuchi.1977.PTPS}),
shell models,
collective models,
relativistic mean-field (RMF) theory
\cite{Ring.1990.AP.v198,Ring.1996.PPNP,Ring_Afanasjev.1997.PPNP,Ring.2001.PPNP,Ring_Serra_Rummel.2001.PPNP,Ring.2007.PPNP,PenaArteaga_Ring.2007.PPNP,Ring.2011.PPNP,Ring.2005.PRep,%
Boguta.1977.NPA.v292,Ring.1974.NPA.v235,Ring.1997.PRC.v55.p540,Ring.1999.ADNDT.v71},
Ab initio calculations theory~\cite{Barrett.2013.PPNP},
QCD approaches for systems of nucleons,
quark models (quark-meson models~\cite{Guichon.2018.PPNP}, potential models)
have been constructed for study of nuclei.
%
In stars, a main focus is given to obtain equation of state of matter (for example, see~\cite{Oertel.2017.RMP}).
Naturally, many from models above are generalized for such a task
(for example, see Refs.~\cite{Danielewicz.2002.Science,Chin.1977.AP,Shen.1998.NPA,Bhuyan.2014.JPG} for RMF theories,
see Ref.~\cite{Oertel.2017.RMP} for Ab initio calculations theory,
etc.).
Today, one of the most accepted models is APR (Akmal, Pandharipande, Ravenhall) model \cite{APR.1998.PRC} constructed in some variants
(see also Refs.~\cite{Pandharipande.1971.NPA,Pandharipande.1971.NPA.v174,Pandharipande.1971.NPA.v166,Pandharipande.1971.NPA.v173}).
It is based on principles of quantum mechanics and can be related with many-nucleon models indicated above.


In this paper we are interesting in different question, that is about ability of nuclear forces to keep nucleons as bound system inside star at high densities of stellar medium.
In particular, it would be interesting to see how such property of nuclear forces is changed in dependence on location of nucleus inside star, and at change of density (gravity) of star.
As it was demonstrated in Ref.~\cite{Maydanyuk.2015.NPA,Maydanyuk_Zhang_Zou.2017.PRC},
even for the same full wave functions with the same boundary conditions there are different nuclear processes (with the same nuclei and energies) where difference in cross-sections of them can reach up to 3-4 times.
RMF theories cannot explain such fully quantum phenomenon, which is not small and important for understanding nuclear processes.
By such a motivation, we would like to use basis of quantum mechanics for analysis in this paper.%
\footnote{There are indications on importance to implement quantum mechanical nuclear models to bremsstrahlung theory in study of emission of photons in stars during proton-capture reactions by nuclei and other nuclear processes
\cite{Maydanyuk_Zhang_Zou.2016.PRC}.}
%
We use compact stars (stars at densities from white dwarf to neutron star \cite{Potekhin.2010.UFN}) in analysis.
It turns out that model of deformed oscillators shells~\cite{Steshenko.1971.YF,Steshenko.1970.UPJ,Steshenko.1971.PhD_thesis,Steshenko.1976.preprint} is enough convenient for such a research
(see straightforward investigations in many nucleon formalism
\cite{Filippov.1980.SJNP,Filippov.1981.SJNP,Vasilevsky.1989.SJNP,Vasilevsky.1990.SJNP,Filippov.1985.SJPN} for basics of the model for the binary cluster configurations for light nuclei,
\cite{Filippov.1985.SJPN,Filippov.1986.SJNP,Filippov.1984.NPA,Filippov.1984.SJNP} for its extensions to describe binary clusters coupled to collective channels,
\cite{Arickx.1996.conf,Vasilevsky.1997.PAN,Filippov.1994.PPN,Vasilevsky.2001.PRC,Vasilevsky.2012.PRC} for three-cluster configurations,
\cite{Lashko_Vasilevsky_Filippov.2019.AnnPhys,Vasilevsky_Kato_Takibayev.2017.PRC,Vasilevsky_Lashko_Filippov.2018.PRC} for newest developments).
After calculations, solution can be obtained in exact analytical form with included additional influence of stellar medium.
%
We apply this simplified model for estimations of emission of bremsstrahlung photons during scattering of protons off nuclei in compact stars.
We study how emission is changed in dependence of characteristics of stellar medium.

The paper is organized by the following way.
In Sec.~\ref{sec.model_nucleus} and \ref{sec.model.star} we present model in estimating binding energy for light nuclei, and influence of stellar medium in frameworks of polytropic model of star.
Analysis of change of binding energy of nuclei in stars is given in Sec.~\ref{sec.analysis}.
In Sec.~\ref{sec.brem.1} we study emission of bremsstrahlung photons during scattering of protons off nuclei from stellar medium of compact stars.
We summarize conclusions in Sect.~\ref{sec.conclusions}.
Derivation of correction of energy of nucleus due to influence of stellar medium is presented in Appendix~\ref{sec.app.1}.

\section{Model of deformed oscillator shells
\label{sec.model_nucleus}}

\subsection{Potential energies of nuclear and Coulomb forces and kinetic energy of nucleus
\label{sec.model.nuclear}}

We define Hamiltonian of system of $A$ nucleons as \cite{Steshenko.1971.YF,Steshenko.1970.UPJ,Steshenko.1976.preprint}
\begin{equation}
\begin{array}{lcl}
  \hat{H}_{DOS} =
  \hat{T} - \hat{T}_{\rm cm} +
  \displaystyle\sum\limits_{i> j =1}^{A} \hat{V}(ij) +
  \displaystyle\sum\limits_{i> j =1}^{A} \displaystyle\frac{e^{2}}{|\vb{r}_{i} - \vb{r}_{j}|}.
\end{array}
\label{eq.model.hamiltonian.1.1}
\end{equation}
We determine potential energy of two-nucleon nuclear forces for nucleus and potential energy of Coulomb forces between protons on the basis of the matrix elements
[see Eqs.~(2.5), (2.6) in ~\cite{Steshenko.1976.preprint}, also Eq.~(15) in~\cite{Steshenko.1971.YF}]:
\begin{equation}
\begin{array}{lcl}
  U_{\rm nucl} =
  \Bigl\langle \Psi(1 \ldots A)
    \Bigl| \sum\limits_{i<j}^{A} \hat{V}_{ij}
    \Bigl| \Psi(1 \ldots A) \Bigr\rangle = \\

  =\; \displaystyle\int
    F_{p} (n; \vb{r}_{1}, \vb{r}_{1})\, F_{p} (n; \vb{r}_{2}, \vb{r}_{2})\,
    \displaystyle\frac{3V_{33}(r_{12}) + V_{13}(r_{12})}{2}\; \vb{dr}_{1}\, \vb{dr}_{2}\; -
  \displaystyle\int
    \bigl| F_{p} (n; \vb{r}_{1}, \vb{r}_{2})\, \bigr|^{2}\,
    \displaystyle\frac{3V_{33}(r_{12}) - V_{13}(r_{12})}{2}\; \vb{dr}_{1}\, \vb{dr}_{2}\; + \\

  +\;
  \displaystyle\int
    F_{p} (n; \vb{r}_{1}, \vb{r}_{1})\, F_{n} (n; \vb{r}_{2}, \vb{r}_{2})\,
    \displaystyle\frac{3V_{33}(r_{12}) + 3V_{31}(r_{12}) + V_{13}(r_{12}) + V_{11}(r_{12})}{2}\; \vb{dr}_{1}\, \vb{dr}_{2}\; - \\
  - \displaystyle\int
    F_{p} (n; \vb{r}_{1}, \vb{r}_{2})\, F_{n} (n; \vb{r}_{2}, \vb{r}_{1})\,
    \displaystyle\frac{3V_{33}(r_{12}) - 3V_{31}(r_{12}) - V_{13}(r_{12}) + V_{11}(r_{12})}{2}\; \vb{dr}_{1}\, \vb{dr}_{2}\; + \\

  +\; \displaystyle\int
    F_{n} (n; \vb{r}_{1}, \vb{r}_{1})\, F_{n} (n; \vb{r}_{2}, \vb{r}_{2})\,
    \displaystyle\frac{3V_{33}(r_{12}) + V_{13}(r_{12})}{2}\; \vb{dr}_{1}\, \vb{dr}_{2}\; -
  \displaystyle\int
    \bigl| F_{n} (n; \vb{r}_{1}, \vb{r}_{2})\, \bigr|^{2}\,
    \displaystyle\frac{3V_{33}(r_{12}) - V_{13}(r_{12})}{2}\; \vb{dr}_{1}\, \vb{dr}_{2},
\end{array}
\label{eq.model.nuclear.1.1}
\end{equation}
and
\begin{equation}
\begin{array}{lcl}
\vspace{1.5mm}
  U_{\rm Coul} & = &
  \Bigl\langle \Psi(1 \ldots A)
    \Bigl| \sum\limits_{i>j=1}^{Z} \displaystyle\frac{e^{2}}{r_{12}}
    \Bigl| \Psi(1 \ldots A) \Bigr\rangle\; = \\
& = &
  2 \displaystyle\int
    F_{p} (n; \vb{r}_{1}, \vb{r}_{1})\, F_{p} (n; \vb{r}_{2}, \vb{r}_{2})\,
    \displaystyle\frac{e^{2}}{r_{12}}\; \vb{dr}_{1}\, \vb{dr}_{2}\; -
  \displaystyle\int
    \bigl| F_{p} (n; \vb{r}_{1}, \vb{r}_{2})\, \bigr|^{2}\,
    \displaystyle\frac{e^{2}}{r_{12}}\; \vb{dr}_{1}\, \vb{dr}_{2},
\end{array}
\label{eq.model.Coulomb.1.1}
\end{equation}
where proton density (for nuclei with even number of protons) is
\begin{equation}
\begin{array}{lcl}
\vspace{1.5mm}
  F_{p} (n; \vb{r}_{i}, \vb{r}_{j}) & = &
  \displaystyle\sum\limits_{s=1}^{z/2}
  \displaystyle\frac
    {\exp\Bigl[ - \displaystyle\frac{1}{2} \Bigl(
      \displaystyle\frac{x_{i}^{2}}{a^{2}} +
      \displaystyle\frac{y_{i}^{2}}{b^{2}} +
      \displaystyle\frac{z_{i}^{2}}{c^{2}} \Bigr)\Bigr]
    \exp\Bigl[ - \displaystyle\frac{1}{2} \Bigl(
      \displaystyle\frac{x_{j}^{2}}{a^{2}} +
      \displaystyle\frac{y_{j}^{2}}{b^{2}} +
      \displaystyle\frac{z_{j}^{2}}{c^{2}} \Bigr)\Bigr]}
   {\pi^{3/2}\, abc\,
   \sqrt{2^{n_{x_{i}}+n_{y_{i}}+n_{z_{i}} + n_{x_{j}}+n_{y_{j}}+n_{z_{j}}}
         n_{x_{i}}! n_{y_{i}}! n_{z_{i}}! n_{x_{j}}! n_{y_{j}}! n_{z_{j}}!}}\; \times \\
  & \times &
   H_{n_{x_{i}}} \Bigl( \displaystyle\frac{x_{i}}{a} \Bigr)
   H_{n_{y_{i}}} \Bigl( \displaystyle\frac{y_{i}}{b} \Bigr)
   H_{n_{z_{i}}} \Bigl( \displaystyle\frac{z_{i}}{c} \Bigr) \cdot
   H_{n_{x_{j}}} \Bigl( \displaystyle\frac{x_{j}}{a} \Bigr)
   H_{n_{y_{j}}} \Bigl( \displaystyle\frac{y_{j}}{b} \Bigr)
   H_{n_{z_{j}}} \Bigl( \displaystyle\frac{z_{j}}{c} \Bigr),
\end{array}
\label{eq.model.density.1.1}
\end{equation}
where summation is performed over all states of needed configuration,
$H_{n}(x)$ are Hermitian polynomials [we use definition from Ref.~\cite{Landau.v3.1989}, p.~749, (а,6)],
$a$, $b$, $c$ are oscillator parameters along axes $x$, $y$, $z$.
Neutron density $F_{n} (n; \vb{r}_{i}, \vb{r}_{j})$ is obtained after change of proton configuration and numbers of states on the corresponding neutron characteristics.
According to definitions (\ref{eq.model.Coulomb.1.1}) and (\ref{eq.model.density.1.1}), proton density is the same for different isotopes. So we obtain the same energy of Coulomb forces for different isotopes.
For two-nucleon potentials we shall use (see Eq.~(1) in Ref.~\cite{Steshenko.1971.YF}):
\begin{equation}
\begin{array}{lcl}
  V_{31}(r) = - 3V_{33}(r) = - V_{t}\, \exp \Bigl( - \displaystyle\frac{r^{2}}{\mu_{t}^{2}} \Bigr), &
  V_{13}(r) = - 1/3 V_{11}(r) = - V_{s}\, \exp \Bigl( - \displaystyle\frac{r^{2}}{\mu_{s}^{2}} \Bigr),
\end{array}
\label{eq.model.potential.1.1}
\end{equation}
where $V_{t} = 72.5$~MeV, $\mu_{t} = 1.47$~fm, $V_{s} = 39,15$~MeV, $\mu_{s} = 1,62$~fm.


We define kinetic energy of system of nucleons (in center-of-mass frame), according to Eq.~(2.4) in Ref.~\cite{Steshenko.1976.preprint} :
%
\begin{equation}
\begin{array}{lcl}
  T_{\rm full} & = &
  \Bigl\langle \Psi(1 \ldots A)
    \Bigl|
      - \displaystyle\frac{\hbar^{2}}{2m} \displaystyle\sum\limits_{i=1}^{A} \laplacian_{i} +
      \displaystyle\frac{\hbar^{2}}{2Am}
        \Bigl( \displaystyle\sum\limits_{i=1}^{A} \grad_{i} \Bigr)^{2}
    \Bigl| \Psi(1 \ldots A) \Bigr\rangle = \\

  & = &
  \displaystyle\frac{A-1}{4}
  \displaystyle\frac{\hbar^{2}}{m}
    \Bigl( \displaystyle\frac{1}{a^{2}} + \displaystyle\frac{1}{b^{2}} + \displaystyle\frac{1}{c^{2}} \Bigr) +
  \displaystyle\frac{\hbar^{2}}{m}
  \biggl\{
    \displaystyle\sum\limits_{s=1}^{Z/2}
    \Bigl( \displaystyle\frac{n_{x,s}}{a^{2}} + \displaystyle\frac{n_{y,s}}{b^{2}} + \displaystyle\frac{n_{z,s}}{c^{2}} \Bigr) +
    \displaystyle\sum\limits_{s^{\prime}=1}^{N/2}
    \Bigl( \displaystyle\frac{n_{x,s^{\prime}}}{a^{2}} + \displaystyle\frac{n_{y,s^{\prime}}}{b^{2}} + \displaystyle\frac{n_{z,s^{\prime}}}{c^{2}} \Bigr)
  \biggr\}.
\end{array}
\label{eq.model.kinetic.1}
\end{equation}

\subsection{Proton and neutron densities
\label{sec.calc.densities.1}}

From Eq.~(\ref{eq.model.density.1.1}) we calculate the proton and neutron densities for isotopes of \isotope[4,6,8]{He} and \isotope[8,10]{Be}:
\begin{equation}
\begin{array}{lll}
  F_{p} (\vb{r}_{1}, \vb{r}_{2}) (\isotope[4]{He}) =
  F_{n} (\vb{r}_{1}, \vb{r}_{2}) (\isotope[4]{He}) = F_{0} (\vb{r}_{1}, \vb{r}_{2}), \\

  F_{p} (\vb{r}_{i}, \vb{r}_{j}) (\isotope[6]{He}) = F_{0} (\vb{r}_{i}, \vb{r}_{j}), \quad
  F_{n} (\vb{r}_{i}, \vb{r}_{j}) (\isotope[6]{He}) = F_{0} (\vb{r}_{i}, \vb{r}_{j}) \cdot \Bigl\{ 1 + \displaystyle\frac{2x_{i}x_{j}}{a^{2}} \Bigr\}, \\

  F_{p} (\vb{r}_{i}, \vb{r}_{j}) (\isotope[8]{He}) = 
  F_{0} (\vb{r}_{i}, \vb{r}_{j}), \quad

  F_{n} (\vb{r}_{i}, \vb{r}_{j}) (\isotope[8]{He}) =
  F_{0} (\vb{r}_{i}, \vb{r}_{j}) \cdot \Bigl\{1 + \displaystyle\frac{2\,x_{i}x_{j}}{a^{2}} + \displaystyle\frac{2\,y_{i}y_{j}}{b^{2}} \Bigr\}, \\

  F_{p} (\vb{r}_{i}, \vb{r}_{j}) (\isotope[8]{Be}) =
  F_{n} (\vb{r}_{i}, \vb{r}_{j}) (\isotope[8]{Be}) =
  F_{0} (\vb{r}_{i}, \vb{r}_{j}) \cdot \Bigl\{ 1 + \displaystyle\frac{2x_{i}x_{j}}{a^{2}} \Bigr\}, \\

  F_{p} (\vb{r}_{i}, \vb{r}_{j}) (\isotope[10]{Be}) =
  F_{p} (\vb{r}_{i}, \vb{r}_{j}) (\isotope[8]{Be}), 

  F_{n} (\vb{r}_{i}, \vb{r}_{j}) (\isotope[10]{Be}) =
  F_{n} (\vb{r}_{i}, \vb{r}_{j}) (\isotope[8]{Be}) + F_{0} (\vb{r}_{i}, \vb{r}_{j}) \cdot \displaystyle\frac{2\,y_{i}y_{j}}{b^{2}},
\end{array}
\label{eq.calc.densities.1.1}
\end{equation}
%
%
%
%
%
where
\begin{equation}
\begin{array}{lcl}
  F_{0} (\vb{r}_{i}, \vb{r}_{j}) & = &
  \displaystyle\frac
    {\exp\Bigl[ - \displaystyle\frac{1}{2} \Bigl(
      \displaystyle\frac{x_{i}^{2}}{a^{2}} +
      \displaystyle\frac{y_{i}^{2}}{b^{2}} +
      \displaystyle\frac{z_{i}^{2}}{c^{2}} \Bigr)\Bigr]
    \exp\Bigl[ - \displaystyle\frac{1}{2} \Bigl(
      \displaystyle\frac{x_{j}^{2}}{a^{2}} +
      \displaystyle\frac{y_{j}^{2}}{b^{2}} +
      \displaystyle\frac{z_{j}^{2}}{c^{2}} \Bigr)\Bigr]}
   {\pi^{3/2}\, abc}.
\end{array}
\label{eq.calc.densities.1.2}
\end{equation}
One can see that the proton and neutron densities are the same for nuclei \isotope[4]{He}, \isotope[8]{Be}, 
they are different for nuclei \isotope[6]{He}, \isotope[8]{He}, \isotope[10]{Be}. 
Also we have properties:
\begin{equation}
  F_{0} (\vb{r}_{1}, \vb{r}_{2}) = F_{0} (\vb{r}_{2}, \vb{r}_{1}) =
  F_{0}^{*} (\vb{r}_{1}, \vb{r}_{2}) = F_{0}^{*} (\vb{r}_{2}, \vb{r}_{1}),
\label{eq.calc.densities.1.3}
\end{equation}
\begin{equation}
\begin{array}{lcl}
  F_{0} (\vb{r}_{1}, \vb{r}_{1}) \cdot F_{0} (\vb{r}_{2}, \vb{r}_{2}) =
  F_{0}^{2} (\vb{r}_{1}, \vb{r}_{2}).
\end{array}
\label{eq.calc.densities.1.4}
\end{equation}
\begin{equation}
\begin{array}{llll}
\vspace{1mm}
  \displaystyle\int F_{0}^{2} (\vb{r}_{1}, \vb{r}_{2}) \cdot
    \exp \Bigl( - \displaystyle\frac{r_{12}^{2}}{\mu^{2}} \Bigr) \cdot \displaystyle\frac{x_{1}^{2}}{a^{2}}\; \vb{dr}_{1} \vb{dr}_{2} =
  \displaystyle\int F_{0}^{2} (\vb{r}_{1}, \vb{r}_{2}) \cdot
    \exp \Bigl( - \displaystyle\frac{r_{12}^{2}}{\mu^{2}} \Bigr) \cdot \displaystyle\frac{x_{2}^{2}}{a^{2}}\; \vb{dr}_{1} \vb{dr}_{2}, \\

  \displaystyle\int F_{0}^{2} (\vb{r}_{1}, \vb{r}_{2}) \cdot
    \exp \Bigl( - \displaystyle\frac{r_{12}^{2}}{\mu^{2}} \Bigr) \cdot \displaystyle\frac{x_{1}^{n}}{a^{n}}\displaystyle\frac{y_{2}^{m}}{b^{m}}\; \vb{dr}_{1} \vb{dr}_{2} =
  \displaystyle\int F_{0}^{2} (\vb{r}_{1}, \vb{r}_{2}) \cdot
    \exp \Bigl( - \displaystyle\frac{r_{12}^{2}}{\mu^{2}} \Bigr) \cdot \displaystyle\frac{x_{2}^{n}}{a^{n}}\displaystyle\frac{y_{1}^{m}}{b^{m}}\; \vb{dr}_{1} \vb{dr}_{2},
\end{array}
\label{eq.calc.densities.1.5}
\end{equation}

\subsection{Potential energy of nuclear two-nucleon interactions
\label{sec.calc.nuclear.1}}

We shall find potential energy of two-nucleon nuclear interactions for nucleus \isotope[6]{He} on the basis of matrix element in Eq.~(\ref{eq.model.nuclear.1.1}) and found densities (\ref{eq.calc.densities.1.1}).
After calculations, we obtain:
\begin{equation}
\begin{array}{llll}
\vspace{0.5mm}
  U_{\rm nucl} (\isotope[4]{He}) = -3\, \bigl\{ V_{t}\, N_{t} + V_{s}\, N_{s} \bigr\}, \\
\vspace{0.5mm}
  U_{\rm nucl} (\isotope[6]{He}) = U_{\rm nucl}^{\rm (sym)} (\isotope[6]{He}) + U_{\rm nucl}^{\rm (asym)} (\isotope[6]{He}), &
    U_{\rm nucl}^{\rm (sym)} (\isotope[6]{He}) = U_{\rm nucl} (\isotope[4]{He}), \\
\vspace{0.5mm}
  U_{\rm nucl} (\isotope[8]{He}) = U_{\rm nucl}^{\rm (sym)} (\isotope[8]{He}) + U_{\rm nucl}^{\rm (asym)} (\isotope[8]{He}), &
    U_{\rm nucl}^{\rm (sym)} (\isotope[8]{He}) = U_{\rm nucl} (\isotope[4]{He}), \\
\vspace{0.5mm}
  U_{\rm nucl} (\isotope[8]{Be}) = U_{\rm nucl}^{\rm (sym)} (\isotope[8]{Be}) + U_{\rm nucl}^{\rm (asym)} (\isotope[8]{Be}), &
  U_{\rm nucl}^{\rm (sym)} (\isotope[8]{Be}) = 2\, U_{\rm nucl} (\isotope[4]{He}), \\
\vspace{0.5mm}
  U_{\rm nucl} (\isotope[10]{Be}) = U_{\rm nucl}^{\rm (sym)} (\isotope[10]{Be}) + U_{\rm nucl}^{\rm (asym)} (\isotope[10]{Be}), &
  U_{\rm nucl}^{\rm (sym)} (\isotope[10]{Be}) = U_{\rm nucl}^{\rm (sym)} (\isotope[8]{Be}), \\
%
\end{array}
\label{eq.calc.nuclear.1.1}
\end{equation}
where
\begin{equation}
\begin{array}{lcl}

\vspace{0.5mm}
  U_{\rm nucl}^{\rm (asym)} (\isotope[6]{He}) =
  -\,
  \Bigl\{
    3\, V_{t}\, N_{t}\, \displaystyle\frac{a_{t}^{2}}{1 + 2\, a_{t}^{2}} +
    V_{s}\,N_{s}\, \displaystyle\frac{\bigl( 1 + 3\, a_{s}^{2} \bigr)\bigl( 1 + 2 a_{s}^{2} \bigr) + 3\, a_{s}^{4}} {\bigl( 1 + 2\, a_{s}^{2} \bigr)^{2}}
  \Bigr\},
\end{array}
\label{eq.calc.nuclear.1.2.6He}
\end{equation}
\begin{equation}
\begin{array}{lll}
\vspace{1.5mm}
  & U_{\rm nucl}^{\rm (asym)} (\isotope[8]{He}) =
  -\,
  \Bigl\{
  V_{t}\, N_{t}\, \Bigl[
    \displaystyle\frac{3\, a_{t}^{2}}{1 + 2\, a_{t}^{2}} + \displaystyle\frac{3\, b_{t}^{2}}{1 + 2\, b_{t}^{2}} -
    \displaystyle\frac{1 + a_{t}^{2}}{1 + 2\, a_{t}^{2}}\, \displaystyle\frac{1 + b_{t}^{2}}{1 + 2\, b_{t}^{2}} +
    \displaystyle\frac{a_{t}^{2}b_{t}^{2}}{(1 + 2\, a_{t}^{2})\, (1 + 2\, b_{t}^{2})}
  \Bigr]\; + \\

  + &
  V_{s}\, N_{s}\, \Bigl[
    \displaystyle\frac{1 + 3\, a_{s}^{2}}{1 + 2\, a_{s}^{2}} +
    \displaystyle\frac{1 + 3\, b_{s}^{2}}{1 + 2\, b_{s}^{2}} +
    \displaystyle\frac{1 + a_{t}^{2}}{1 + 2\, a_{t}^{2}}\, \displaystyle\frac{1 + b_{t}^{2}}{1 + 2\, b_{t}^{2}} +
    \displaystyle\frac{a_{s}^{2}b_{s}^{2}}{(1 + 2\, a_{s}^{2})\, (1 + 2\, b_{s}^{2})} +
    \displaystyle\frac{3\, a_{s}^{4}}{( 1 + 2\, a_{s}^{2} )^{2}} +
    \displaystyle\frac{3\, b_{s}^{4}}{( 1 + 2\, b_{s}^{2} )^{2}}
  \Bigr]\, \Bigr\},
\end{array}
\label{eq.calc.nuclear.1.2.8He}
\end{equation}
\begin{equation}
\begin{array}{lcl}
  U_{\rm nucl}^{\rm (asym)} (\isotope[8]{Be}) =
  - 9\;
  \Bigl\{
    V_{t}\, N_{t}\, \displaystyle\frac{a_{t}^{4}} {( 1 + 2 a_{t}^{2} )^{2}} +
    V_{s}\, N_{s}\, \displaystyle\frac{a_{s}^{4}} {( 1 + 2 a_{s}^{2} )^{2}}
  \Bigr\},
\end{array}
\label{eq.calc.nuclear.1.2.8Be}
\end{equation}
\begin{equation}
\begin{array}{lll}
\vspace{0.7mm}
  & U_{\rm nucl}^{\rm (asym)} (\isotope[10]{Be}) = U_{\rm nucl}^{\rm (asym)} (\isotope[8]{Be})\; - \\
  - & \Bigl\{ 3\, N_{t}\, \displaystyle\frac{b_{t}^{2}}{1 + 2\, b_{t}^{2}} + 4\, N_{s}\, \displaystyle\frac{b_{s}^{2}}{1 + 2\, b_{s}^{2}}
  + 3\, N_{s}\, \displaystyle\frac{b_{s}^{4}}{( 1 + 2\, b_{s}^{2})^{2}} \Bigr\} -
  3 \cdot \Bigl\{
      \displaystyle\frac{N_{t}\, a_{t}^{2}b_{t}^{2}}{(1 + 2\, a_{t}^{2})\, (1 + 2\, b_{t}^{2})} +
      \displaystyle\frac{N_{s}\, a_{s}^{2}b_{s}^{2}}{(1 + 2\, a_{s}^{2})\, (1 + 2\, b_{s}^{2})}
    \Bigr\},
\end{array}
\label{eq.calc.nuclear.1.2.10Be}
\end{equation}
and we use notations (with change of indexes $t$ and $s$):
\begin{equation}
\begin{array}{llllllll}
  a_{t} = a/\mu_{t}, & b_{t} = b/\mu_{t}, & c_{t} = c/\mu_{t}, &
  a_{s} = a/\mu_{s}, & b_{s} = b/\mu_{s}, & c_{s} = c/\mu_{s},
\end{array}
\label{eq.calc.nuclear.1.3}
\end{equation}
\begin{equation}
  N_{t} = \displaystyle\frac{1}{\sqrt{1 + 2a_{t}^{2}} \sqrt{1 + 2b_{t}^{2}} \sqrt{1 + 2c_{t}^{2}}}.
\label{eq.calc.nuclear.1.4}
\end{equation}
From solutions above one can see that
(1) only \isotope[4]{He} is spherical in the ground state, while other nuclei are deformed;
(2) nuclei \isotope[6]{He},  \isotope[8]{Be} are axially symmetric in the ground state,
while \isotope[8]{He},  \isotope[10]{Be} are fully deformed.
Further calculations of minima of full energy of these nuclei confirm such a logic.

In spherically symmetric approximation ($a = b = c$), we obtain
\begin{equation}
\begin{array}{lcl}
\vspace{1mm}
  U_{\rm nucl} (\isotope[6]{He}) =
  -\, 3\,V_{t}\, N_{t}^{\rm (sph)}\: \Bigl\{ 1 + \displaystyle\frac{a_{t}^{2}}{1 + 2\, a_{t}^{2}} \Bigr\} -
  V_{s}\, N_{s}^{\rm (sph)}\, \biggl\{ 3 + \displaystyle\frac{\bigl( 1 + 3\, a_{s}^{2} \bigr) \bigl( 1 + 2 a_{s}^{2} \bigr) + 3\, a_{s}^{4}} {\bigl( 1 + 2\, a_{s}^{2} \bigr)^{2}} \biggr\}, \\
%
\vspace{1mm}
  U_{\rm nucl} (\isotope[8]{He}) = U_{\rm nucl} (\isotope[4]{He}) -
  \Bigl\{
    V_{t}\, N_{t}^{\rm (sph)}\, \Bigl[ \displaystyle\frac{6\, a_{t}^{2} - 1}{1 + 2\, a_{t}^{2}} \Bigr] +
    V_{s}\, N_{s}^{\rm (sph)}\, \Bigl[ 3 + \displaystyle\frac{8\, a_{s}^{4}}{( 1 + 2\, a_{s}^{2} )^{2}} \Bigr]\,
  \Bigr\}, \\
%
\vspace{1mm}
  U_{\rm nucl} (\isotope[8]{Be}) =
  - 3\,
  \Bigl\{
    V_{t}\, N_{t}^{\rm (sph)}\, \Bigl[2 + \displaystyle\frac{3\,a_{t}^{4}} {( 1 + 2 a_{t}^{2} )^{2}} \Bigr] +
    V_{s}\, N_{s}^{\rm (sph)}\, \Bigl[2 + \displaystyle\frac{3\,a_{s}^{4}} {( 1 + 2 a_{s}^{2} )^{2}} \Bigr]
  \Bigr\}, \\
%
  U_{\rm nucl} (\isotope[10]{Be}) = U_{\rm nucl} (\isotope[8]{Be}) -
  \Bigl\{ 3\, N_{t}\, \displaystyle\frac{a_{t}^{2}}{1 + 2\, a_{t}^{2}} +
    3\, N_{t}\, \displaystyle\frac{a_{t}^{4}}{(1 + 2\, a_{t}^{2})^{2}} +
    4\, N_{s}\, \displaystyle\frac{a_{s}^{2}}{1 + 2\, a_{s}^{2}} +
    6\, N_{s}\, \displaystyle\frac{a_{s}^{4}}{( 1 + 2\, a_{s}^{2})^{2}}
  \Bigr\}.
\end{array}
\label{eq.calc.nuclear.1.5}
\end{equation}
where
\begin{equation}
  N_{t}^{\rm (sph)} = (1 + 2a_{t}^{2})^{-3/2}.
\label{eq.calc.nuclear.1.6}
\end{equation}

\subsection{Kinetic energy of nuclei
\label{sec.calc.kinetic.1.1}}

We calculate kinetic energy of nuclei, according to Eq.~(\ref{eq.model.kinetic.1}), and obtain:
\begin{equation}
\begin{array}{lcl}
\vspace{1mm}
  T_{\rm full} (\isotope[4]{He}) & = &
  \displaystyle\frac{3}{4}
  \displaystyle\frac{\hbar^{2}}{m}
    \Bigl(
      \displaystyle\frac{1}{a^{2}} +
      \displaystyle\frac{1}{b^{2}} +
      \displaystyle\frac{1}{c^{2}}
    \Bigr), \\

\vspace{1mm}
  T_{\rm full} (\isotope[6]{He}) & = &
  \displaystyle\frac{5}{3} \cdot T_{\rm full} (\isotope[4]{He}) + \displaystyle\frac{\hbar^{2}}{m\,a^{2}}, \\

\vspace{1mm}
  T_{\rm full} (\isotope[8]{He}) & = &
  \displaystyle\frac{7}{3} \cdot T_{\rm full} (\isotope[4]{He}) +
  \displaystyle\frac{\hbar^{2}}{m} \Bigl( \displaystyle\frac{1}{a^{2}} + \displaystyle\frac{1}{b^{2}} \Bigr), \\

\vspace{1mm}
  T_{\rm full} (\isotope[8]{Be}) & = &
  \displaystyle\frac{7}{3} \cdot T_{\rm full} (\isotope[4]{He}) + 2\, \displaystyle\frac{\hbar^{2}}{m\,a^{2}} =
  T_{\rm full} (\isotope[8]{He}) + \displaystyle\frac{\hbar^{2}}{m} \Bigl( \displaystyle\frac{1}{a^{2}} - \displaystyle\frac{1}{b^{2}} \Bigr), \\

  T_{\rm full} (\isotope[10]{Be}) & = &
  \displaystyle\frac{9}{3} \cdot T_{\rm full} (\isotope[4]{He}) + \displaystyle\frac{\hbar^{2}}{m} \Bigl( \displaystyle\frac{2}{a^{2}} + \displaystyle\frac{1}{b^{2}} \Bigr).
\end{array}
\label{eq.calc.kinetic.1.1}
\end{equation}
In the spherically symmetric case ($a = b = c$) we obtain:
\begin{equation}
\begin{array}{lllllllll}
\vspace{1.7mm}
  T_{\rm full}^{\rm sph} (\isotope[4]{He}) = \displaystyle\frac{9}{4} \displaystyle\frac{\hbar^{2}}{m\, a^{2}}, &

  T_{\rm full} (\isotope[6]{He}) =
  \displaystyle\frac{19}{9} \cdot T_{\rm full} (\isotope[4]{He}), &

  T_{\rm full} (\isotope[8]{He}) =
  \displaystyle\frac{29}{9} \cdot T_{\rm full} (\isotope[4]{He}), \\

  T_{\rm full} (\isotope[8]{Be}) = T_{\rm full} (\isotope[8]{He}), &

  T_{\rm full} (\isotope[10]{Be}) =
  \displaystyle\frac{31}{9}\: T_{\rm full} (\isotope[4]{He}) =
  \displaystyle\frac{31}{29}\, T_{\rm full} (\isotope[8]{Be}).
\end{array}
\label{eq.calc.kinetic.1.2}
\end{equation}

\section{Energy of nucleus inside compact stars
\label{sec.model.star}}

\subsection{Polytropic stars
\label{sec.star.1}}


Star without rotation and magnetic field has spherical form.
Its equilibrium is determined by balance of forces of gravity and gradient of pressure.
Nuclear reactions take place in stars and there is radiation from their surfaces.
For relativistic stars on last stage of evolution, pressure $P$ depends on density $\rho$ only and can be described by equation of state of $P = P(\rho)$.
At some approximation, star under conditions above can be described by Lane-Emden equation (see Ref.~\cite{Bisnovatyi-Kogan.2011.book}, p.~19):
\begin{equation}
\begin{array}{lcl}
  \displaystyle\frac{d}{d\xi}
  \Bigl(
    \xi^{2}\,
    \displaystyle\frac{d\theta}{d\xi}
  \Bigr) = -\xi^{2}\, \theta^{n}
\end{array}
\label{eq.star.1.1}
\end{equation}
with boundary conditions of $\theta(0) = 1$, $d\theta(0) / d\xi = 0$.
Here, $\xi$ is a dimensionless distance from center of star and $\theta$ is related density.
$n$ is the polytropic index that appears in the polytropic equation of state:
\begin{equation}
\begin{array}{llll}
  P = K \cdot \rho^{\gamma}, &
  \gamma = 1 + \displaystyle\frac{1}{n},
\end{array}
\label{eq.star.1.2}
\end{equation}
where $P$ and $\rho$ are the pressure and density,
$K$ is constant of proportionality.
Density $\rho(r)$ inside star at distance $r$ from center of star is derived as
\begin{equation}
\begin{array}{llllll}
  \rho (r) = \rho_{\rm c} \cdot \theta^{n}, &
  r = R_{0} \cdot \xi, &
  R_{0}^{2} = \displaystyle\frac{(n+1)\,K}{4\pi\, G} \rho_{c}^{\frac{1}{n} - 1},
\end{array}
\label{eq.star.1.3}
\end{equation}
where $\rho_{\rm c}$ is pressure at center of star, $R_{0}$ is parameter.
In frameworks of such a model, radius and mass of star are calculated as (see Ref.~\cite{Bisnovatyi-Kogan.2011.book}, p.~22):
\begin{equation}
\begin{array}{llllll}
  r_{R} = \xi_{R} \cdot R_{0}, &
  M = \displaystyle\int\limits_{0}^{r_{R}} 4\pi\rho\; r^{2}dr =
  4\pi\, \Bigl[ \displaystyle\frac{(n+1)\,K}{4\pi\, G} \Bigr]^{3/2}\, \rho_{c}^{\displaystyle\frac{3-n}{2n}}\, \displaystyle\int\limits_{0}^{\xi_{1}} \theta^{n} \xi^{2}\, d\xi.
\end{array}
\label{eq.star.1.4}
\end{equation}
$\xi_{R}$ is dimensionless radius of star defined from condition $\theta (\xi_{R})= 0$.
In particular, mass of star at $n=3$ does not depend on density $\rho_{c}$. 

\vspace{0.8mm}
In this paper we will be interesting in what happens with nucleus in dependence on depth of its location inside star of such a type.
Step-by-step, we will change a distance from center of star to this nucleus and analyze how much strong are forces keeping nucleons of this nucleus as bound quantum system.
Clear understanding can be obtained from binding energy of  nucleus as system of nucleons. We calculate binding energy as summation of potential energy of nuclear forces, Coulomb forces, kinetic energy of nucleons described above.
It turns out, that the simplest case of $n=3$ allows to obtain a clear picture (other cases add more technical derivations, so we will omit them in this paper).

\subsection{Quantum mechanics of nucleus under influence of stelar medium
\label{sec.star.2}}

Let us write full hamiltonian of nucleus with additional influence of medium of star on nucleons of this nucleus as
%
\begin{equation}
\begin{array}{lcl}
  \hat{H} =
  - \displaystyle\frac{\hbar^{2}}{2m} \displaystyle\sum\limits_{i=1}^{A} \grad_{i}^{2} +
  \displaystyle\sum\limits_{i,j=1}^{A} V_{\rm DOS} (\abs{\vb{r}_{i} - \vb{r}_{j}}) +
  \displaystyle\sum\limits_{i,j=1}^{A} V_{\rm star} (\vb{r}_{i}, \vb{r}_{j}).
\end{array}
\label{eq.star.2.1}
\end{equation}
In the first approximation, we shall assume that influence of stellar medium on nucleons of the studied nucleus is homogeneous.
Force $\vb{F}$ of such an influence should depend on distance $R$ between center of star and center of mass of the studied nucleus.
Potential of such an influence should depend on relative distances between nucleons of the studied nucleus.%
%
\footnote{Gradient of potential $U$ with opposite sign is force $\vb{F}_{P}$ acting on particle with mass $m$: $\vb{F}_{P} \equiv - \grad U$.
One can clarify that, if to analyze action on particle in quantum mechanics in semiclassical approximation (see Ref.~\cite{Landau.v3.1989}, p.~209).
Operator of velocity in quantum mechanics as $m\, \vu{\dot{v}} = - \grad U$ (see~(19.3), Ref.~\cite{Landau.v3.1989}, p.~82)
indicates on such a relation between $U$ and $\vb{F}_{P}$ also.
From here, one can obtain potential. In particular, for homogeneous force one can find:
\begin{equation}
  U (\vb{r}) = - \vb{F}_{P}\, \displaystyle\int\limits_{}^{} \vb{dr} = - \vb{F}_{P}\, \vb{r}.
\label{eq.star.3.1}
\end{equation}
%
}
%
Such a formalism is given in quantum mechanics (see Ref.~\cite{Landau.v3.1989}, p.~100--102, for details), therefore, we define it as%
\footnote{In order to understand, which sign should be used in this formula, we return back to logics in Ref.~\cite{Landau.v3.1989} (see p.~100 in this book).
In particular, at increasing of distances between nucleons, $|\vb{r}_{i} - \vb{r}_{j}|$, potential of influence stelar medium should suppress (i.e. not reinforce) relative leaving of nucleons from the nucleus.
Therefore, the potential should increase (not decrease) at increasing of $|\vb{r}_{i} - \vb{r}_{j}|$.
I.e. sign if Eq.~(\ref{eq.star.2.2}) is chosen correctly.}
%
\begin{equation}
\begin{array}{lcl}
  V_{\rm star} (R, \vb{r}_{i}, \vb{r}_{j}) = +\, \abs{\vb{F}_{P} (R) \cdot (\vb{r}_{i} - \vb{r}_{j})}.
\end{array}
\label{eq.star.2.2}
\end{equation}
Corresponding correction $\Delta E_{\rm star}$ to the full energy of nucleus due to inclusion of influence of star on nucleons of nucleus can be defined as
\begin{equation}
\begin{array}{lcl}
  \Delta E_{\rm star} & = &
  \Bigl\langle \Psi(1 \ldots A)
    \Bigl|
      \displaystyle\sum\limits_{i,j=1}^{A} V_{\rm star} (R, \vb{r}_{i}, \vb{r}_{j})
    \Bigl| \Psi(1 \ldots A) \Bigr\rangle.
\end{array}
\label{eq.star.2.3}
\end{equation}
Calculations of such a matrix element are presented in Appendix~\ref{sec.app.1}. For even-even nuclei (at $Z=N$) we obtain:
\begin{equation}
\begin{array}{ll}
\vspace{0.9mm}
  & \langle \Psi (1 \ldots A)\, |\, \hat{V}\, (\vb{r}_{i}, \vb{r}_{j}) |\, \Psi (1 \ldots A) \rangle =

  \displaystyle\frac{1}{A \cdot (A-1)}\;
  \displaystyle\sum\limits_{k=1}^{A}
  \displaystyle\sum\limits_{m=1, m \ne k}^{A}\;
    \Bigl\langle
      \varphi_{0} (\vb{r}_{i})\, \varphi_{0} (\vb{r}_{j}) \Bigl|\, \hat{V}\, (\vb{r}_{i}, \vb{r}_{j}) \Bigr|\, \varphi_{0} (\vb{r}_{i})\, \varphi_{0} (\vb{r}_{j})
    \Bigr\rangle\; = \\

  = &
    \Bigl\langle
      \varphi_{0} (\vb{r}_{i})\, \varphi_{0} (\vb{r}_{j}) \Bigl|\, \hat{V}\, (\vb{r}_{i}, \vb{r}_{j}) \Bigr|\, \varphi_{0} (\vb{r}_{i})\, \varphi_{0} (\vb{r}_{j})
    \Bigr\rangle.
\end{array}
\label{eq.star.2.4}
\end{equation}
In particular, for \isotope[4]{He} we have (see Eq.~(\ref{eq.app.1.11}) at $a=b=c$)
\begin{equation}
\begin{array}{lcl}
  \Delta E_{\rm star} (\isotope[4]{He}) & = &
  12 \cdot \vb{F}_{P} (R) \cdot \displaystyle\int F_{0}^{2} (\vb{r}_{1}, \vb{r}_{2})\, (\vb{r}_{1} - \vb{r}_{2})\; \vb{dr}_{1}\, \vb{dr}_{2} =
  \displaystyle\frac{12 \cdot 2^{3/2}\, a}{\pi^{1/2}} \cdot F_{P} (R).
\end{array}
\label{eq.star.2.5}
\end{equation}
where $F_{P} (R) = |\vb{F}_{P} (R)|$.
Even without numerical estimations, now picture of influence of stellar medium on the studied nucleus has became clear.
Forces of stellar medium press on nucleons of nucleus.
The deeper this nucleus is located in star, the stronger such forces press on nucleus.
However, binding energy (it is negative for nucleus in the external layer of star) is increased at deeper location of this nucleus in star.
Starting from some critical distance from nucleus to center of star, the binding energy becomes positive.
This means that full energy of individual nucleons of the studied nucleus is already larger than mass of this nucleus,
i.e. we obtain unbound system of nucleons and nucleus is disintegrated on nucleons.
Now it could be interesting to estimate if such a phenomenon is appeared in white dwarfs and neutron stars in frameworks of the model above.
%
The kinetic energy is increased at deeper location of nucleus in star.
At decreasing distance from the studied nucleus to center of star, change of kinetic energy is unlimited, while change of nuclear energy is limited.
So, ratio between kinetic energy of nucleons of nucleus and nuclear energy of nucleus is changed also.

%

\section{Analysis
\label{sec.analysis}}

One of objects, where polytropic model is successfully applied, is wight dwarf
(see indications in Ref.~\cite{Bisnovatyi-Kogan.1989.book}, p.~364--370;
Ref.~\cite{Kippenhahn.2012.book}, p.~213--233, p.~475--496).
Thus, according to Ref.~\cite{Bisnovatyi-Kogan.2011.book} (see Fig.~2.2 in that book, p.~33; also Fig.~103 in Ref.~\cite{Bisnovatyi-Kogan.1989.book}, p.~365),
density in center of such a star is in the region of $10^{+5} {\rm g}\, {\rm cm}^{-3}$ -- $1.4 \cdot 10^{+9}\, {\rm g}\, {\rm cm}^{-3}$ .
So, let us start analysis from such a type of star.

\begin{figure}[htbp]
\centerline{\includegraphics[width=90mm]{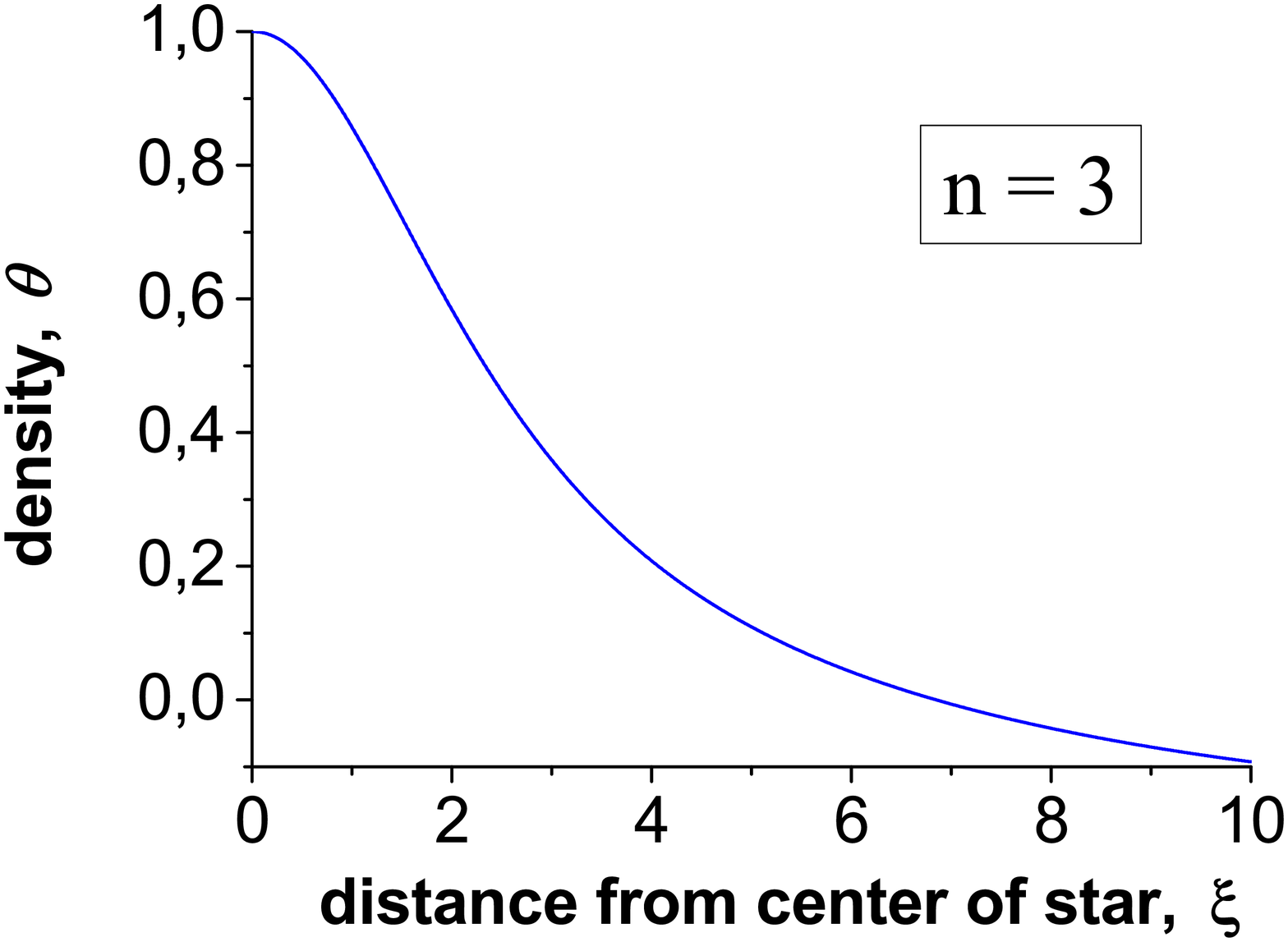}
\hspace{-1mm}\includegraphics[width=88mm]{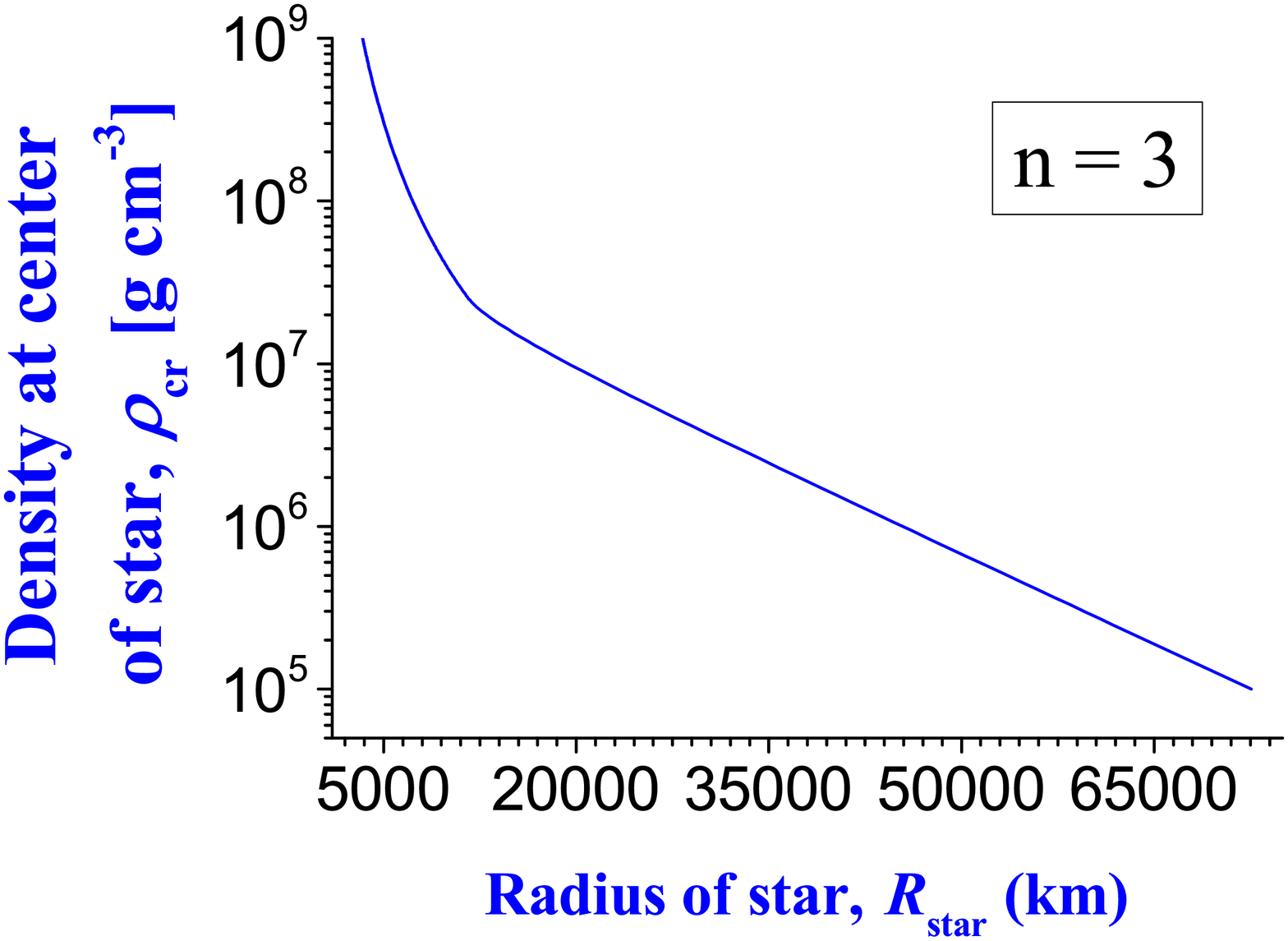}}
\vspace{-3mm}
\caption{\small (Color online)
Panel (a): Solution of Lane-Emden equation (\ref{eq.star.1.1}) at $n=3$ by the finite-difference method
[parameters of calculations:
boundary conditions are $\theta(0) = 1$, $d\theta(0) / d\xi = 0$
].
In figure one can see monotonous decreasing of density $\theta$ at increasing of distance $\xi$.
According to the model, internal region of starcorresponds to condition $\theta (\xi) \ge 0$,
and radius of star, $\xi_{\rm r}$, is found from condition of $\theta (\xi_{\rm r}) = 0$.
Panel (b): Radius of star in dependence on density in its center (densities are chosen for white dwarfs, at $n=3$).
\label{fig.1}}
\end{figure}

At first, let us see how density is dependent on distance to center of star in frameworks of such a model.
In Fig.~\ref{fig.1}~(a) one cab see solution of Lane-Emden equation (\ref{eq.star.1.1}) at $n=3$ by the finite-difference method.
Radius of star is determined from condition of $\theta (\xi_{\rm r}) = 0$, we obtain $\xi_{\rm r} = 6.881$.
%
%
In Fig.~\ref{fig.1}~(b) one can see radius of star in dependence on its density at center for such a model.
One can see that such a model gives white dwarfs with radiuses in region
from 3011.28~kilometres (at $\rho_{\rm cr} = 1.4 \cdot 10^{9}\, {\rm g}\, {\rm cm}^{-3}$)
to 72~576.27~kilometres (at $\rho_{\rm cr} = 10^{5}\, {\rm g}\, {\rm cm}^{-3}$).

\begin{figure}[htbp]
\centerline{\includegraphics[width=90mm]{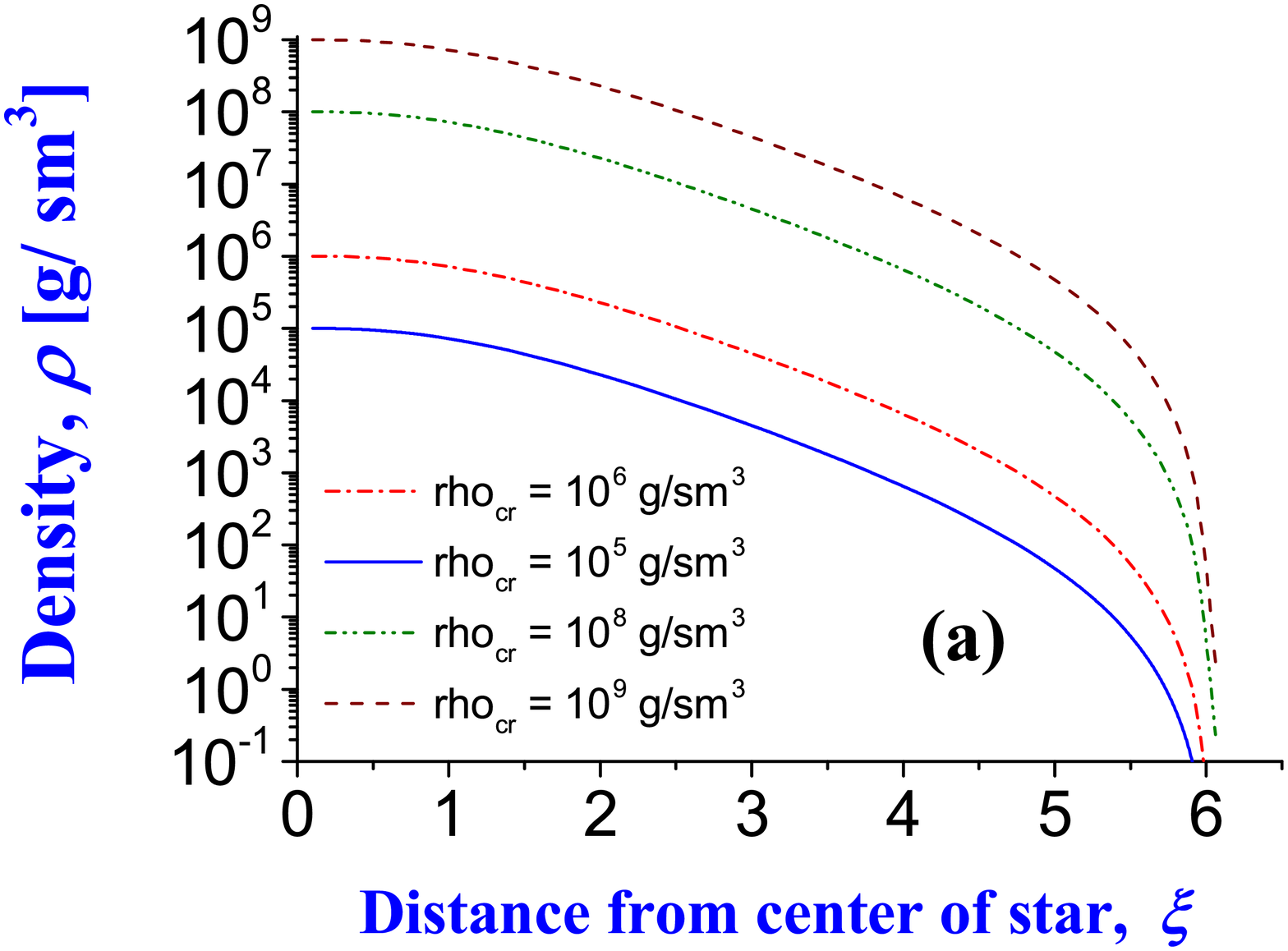}
\hspace{-1mm}\includegraphics[width=90mm]{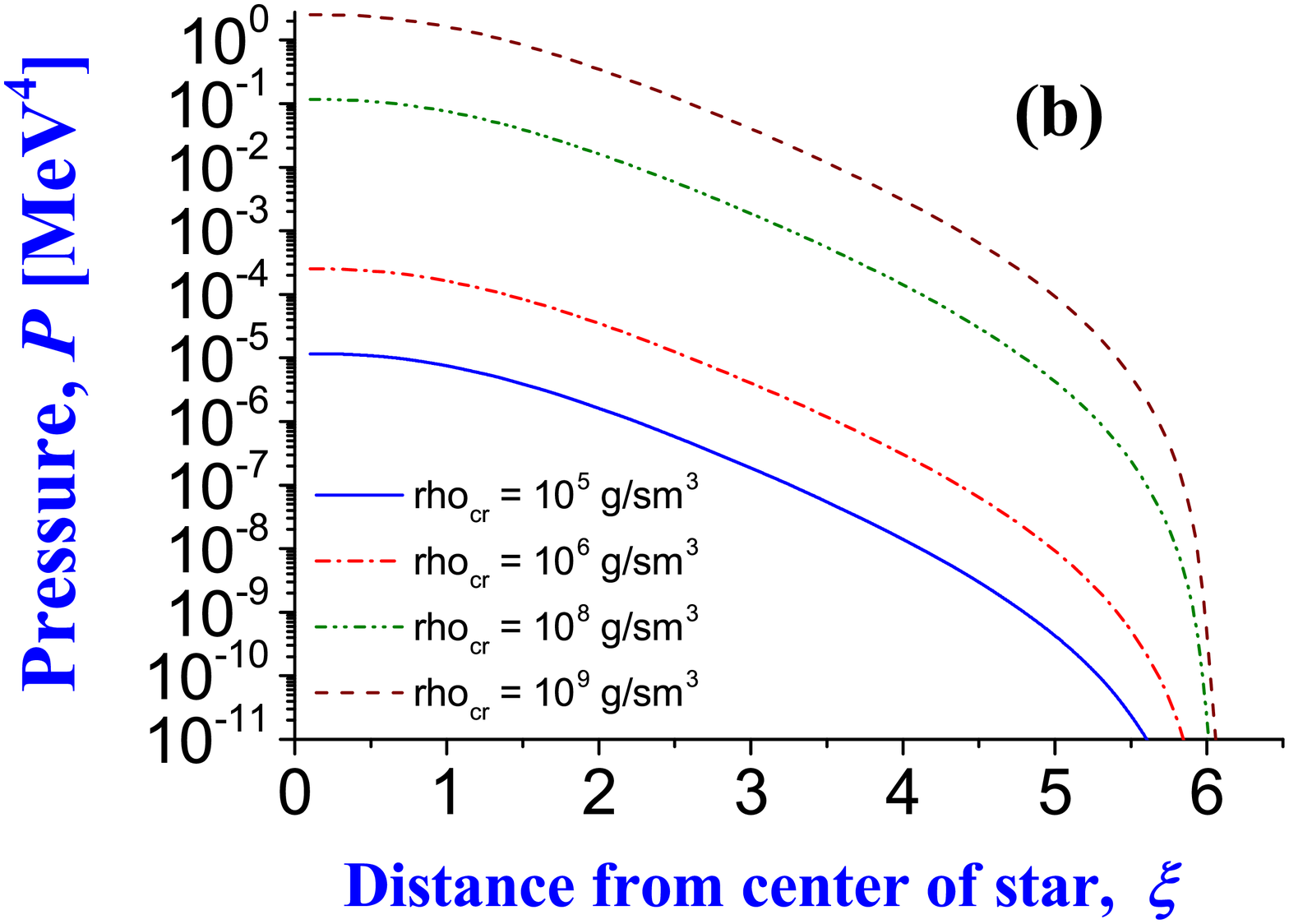}}
\vspace{-2mm}
\caption{\small (Color online)
Density $\rho$ (a) and pressure $P$ (b) inside star in dependence on distance $\xi$ from center of star at $n=3$
[parameters of calculations:
density is defined in Eq.~(\ref{eq.star.1.3}),
pressure is defined in Eq.~(\ref{eq.star.1.2})
].
\label{fig.2}}
\end{figure}

We shall analyze, how the density $\rho$ in star is changed in dependence on distance from center of star.
Results of such calculations at $n=3$ are presented in Fig.~\ref{fig.2}~(a).
As nest step, we shall estimate pressure in star in dependence of distance between nucleus (its center of mass) and center of mass of star.
Using Eqs.~(\ref{eq.star.1.2}), at $n=3$ we have (see Eq.~(2.3), p.~32 in Ref.~\cite{Bisnovatyi-Kogan.2011.book}):
\begin{equation}
\begin{array}{lll}
  \gamma = \displaystyle\frac{4}{3}, &
  P_{n=3} = K \cdot \rho^{4/3}, &
  K_{n=3} = \displaystyle\frac{(3\, \pi^{2})^{1/3}}{4}\, \hbar c\, (\mu_{e}\, m_{\rm p})^{-4/3} = 3.384\, 782 \cdot 10^{-5}\, \mbox{MeV}^{-4/3}.
\end{array}
\label{eq.star.1.3.4}
\end{equation}
Results of such calculations are presented in Fig.~\ref{fig.2}~(b).

\begin{figure}[htbp]
\centerline{\includegraphics[width=90mm]{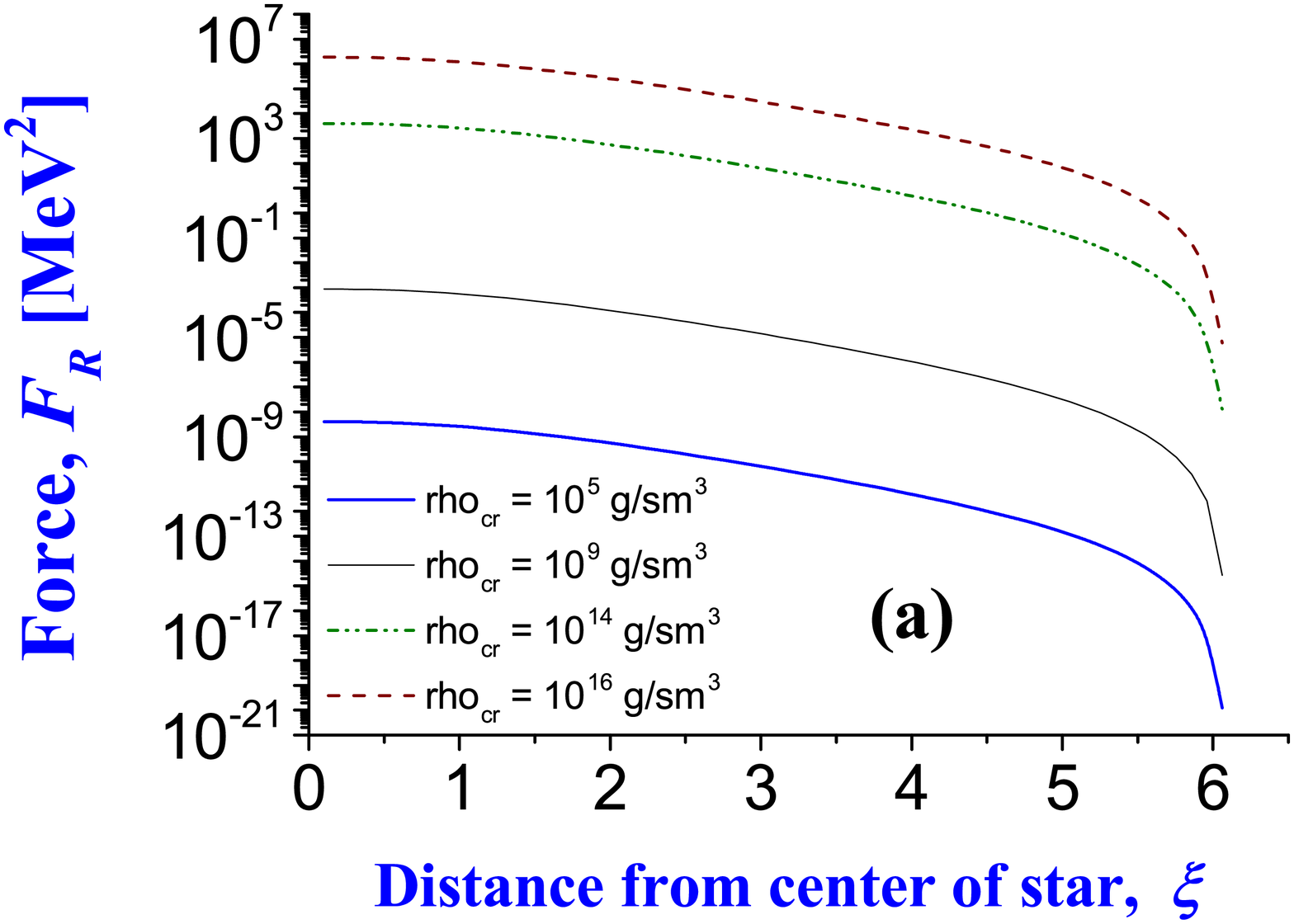}
\hspace{-1mm}\includegraphics[width=90mm]{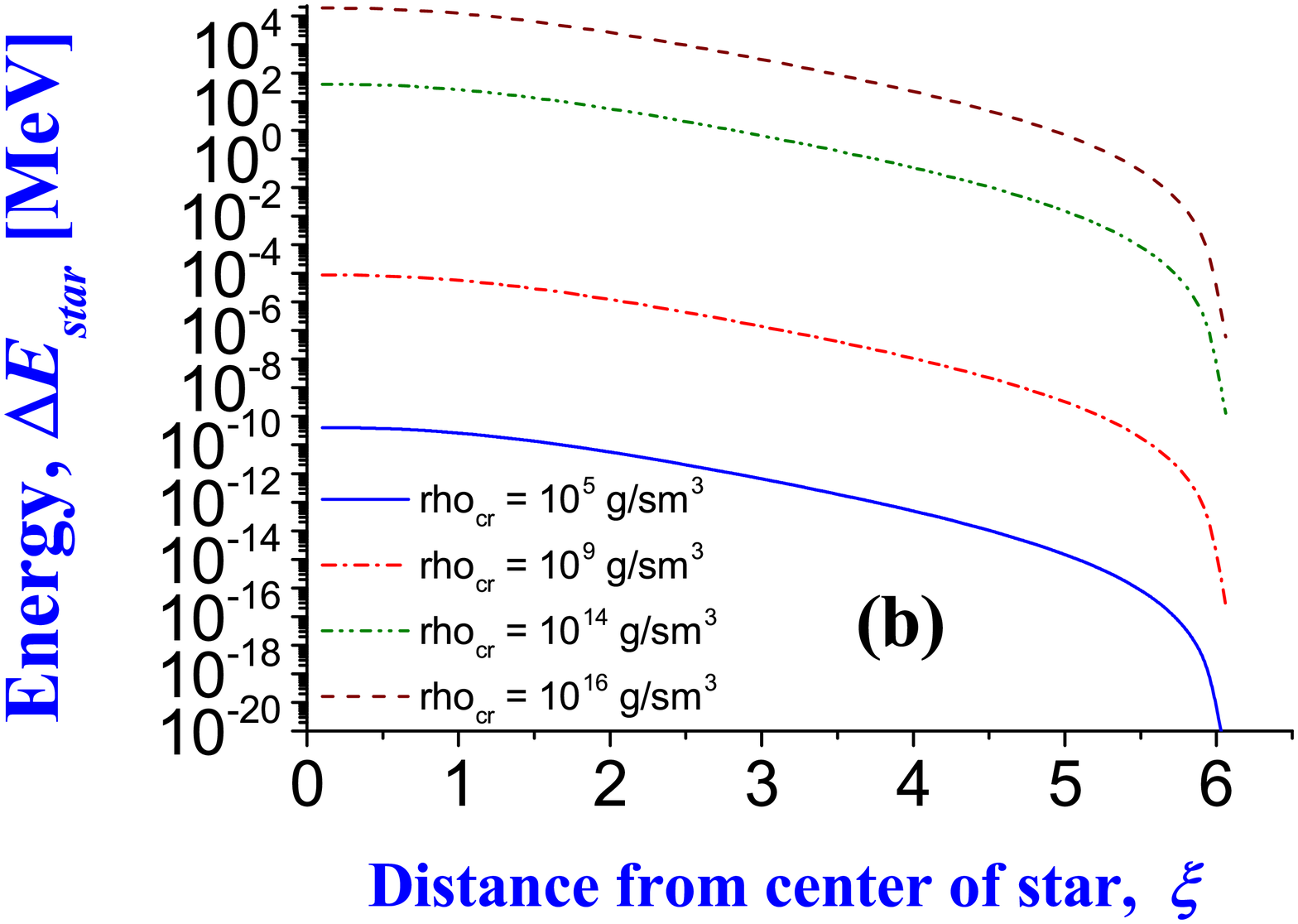}}
\vspace{-2mm}
\caption{\small (Color online)
Panel (a): Force $F_{R}$, acting on nucleons of nucleus \isotope[4]{He}, in dependence on its distance $\xi$ to the center of star (at $n=3$)
[Force is defined in Eq.~(\ref{eq.star.1.3.1}), oscillator parameter $a=1.05$~fm is fixed for estimations, that is close to minimum of full energy of \isotope[4]{He} in natural conditions (in Earth)].
Panel (b): Correction $\Delta E$ to energy of nucleus \isotope[4]{He}, in result of influence of stellar medium , in dependence on distance to center of star (at $n=3$)
[Correction of energy $\Delta E$ is defined in Eq.~(\ref{eq.star.2.5})].
\label{fig.3}}
\end{figure}

Now let us estimate force acting on the studied nucleus in star in result of influence of stellar medium.
We shall find it on the basis of pressure stellar medium.
According to definition, pressure is force applied perpendicular to the surface of an object per unit area over which that force is distributed.
So, pressure acting on selected layer inside star, represents force, acting on unit area of this layer.
Then, force acting on full such a layer can be found as the pressure multiplied on the full area of this layer.
Let us consider nucleus with surface $S_{\rm nucl}$ inside star. One can determine force acting on this nucleus as pressure multiplied on area of surface of this nucleus as
%
%
\begin{equation}
\begin{array}{lll}
  F_{R} (R) = P(R) \cdot S_{\rm nucl}.
\end{array}
\label{eq.star.1.3.1}
\end{equation}
In the simplest approximation, the studied nucleus can be considered in the spherical form where its area is $S_{\rm nucl} = 4\pi\, R_{\rm nucl}^{2}$, $R_{\rm nucl} = a \cdot A^{1/3}$.
Results of calculations of such a force acting on nucleus \isotope[4]{He} in star are presented in Fig.~\ref{fig.3}~(a), if pressure is shown in Fig.~\ref{fig.2}~(b)%
\footnote{For simplicity of presentation, in this paper we use units for force in MeV (that is used in computer calculations and allows to study physical process inside distances of nuclei and stars, at the same time).
This do not forbid to perform comparable analysis.
}.

After obtaining force, we shall find correction to full energy of nucleus from such an influence.
Results of such calculations for nucleus \isotope[4]{He} are presented in Fig.~\ref{fig.3}~(b).
From this figure one can see that in the white dwarfs (corresponding to densities in region of
$10^{5}\, {\rm g}\, {\rm cm}^{-3}$ -- $1.4 \cdot 10^{9}\, {\rm g}\, {\rm cm}^{-3}$) nucleus cannot be disintegrated (in this model).
However, we see that for more high densities this phenomenon really happens, starting from some critical distances from center of stars
[see upper brown dashed line (at $\rho_{\rm cr} = 10^{16}\, {\rm g}\, {\rm cm}^{-3}$) and
green dash-double dotted line (at $\rho_{\rm cr} = 10^{14}\, {\rm g}\, {\rm cm}^{-3}$) in figure]. This case corresponds to the neutron stars.



\begin{figure}[htbp]
\centerline{\includegraphics[width=90mm]{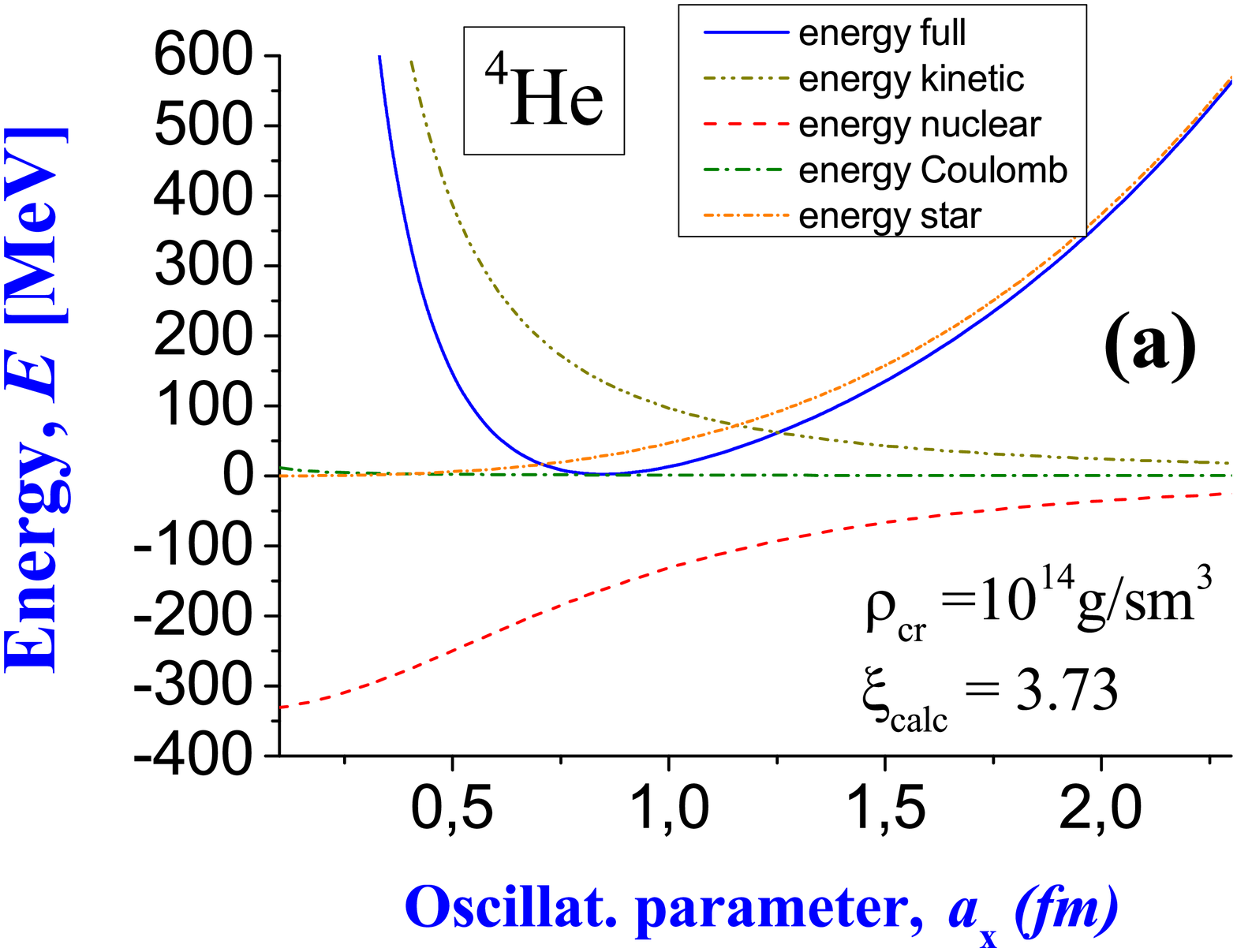}
\hspace{-1mm}\includegraphics[width=90mm]{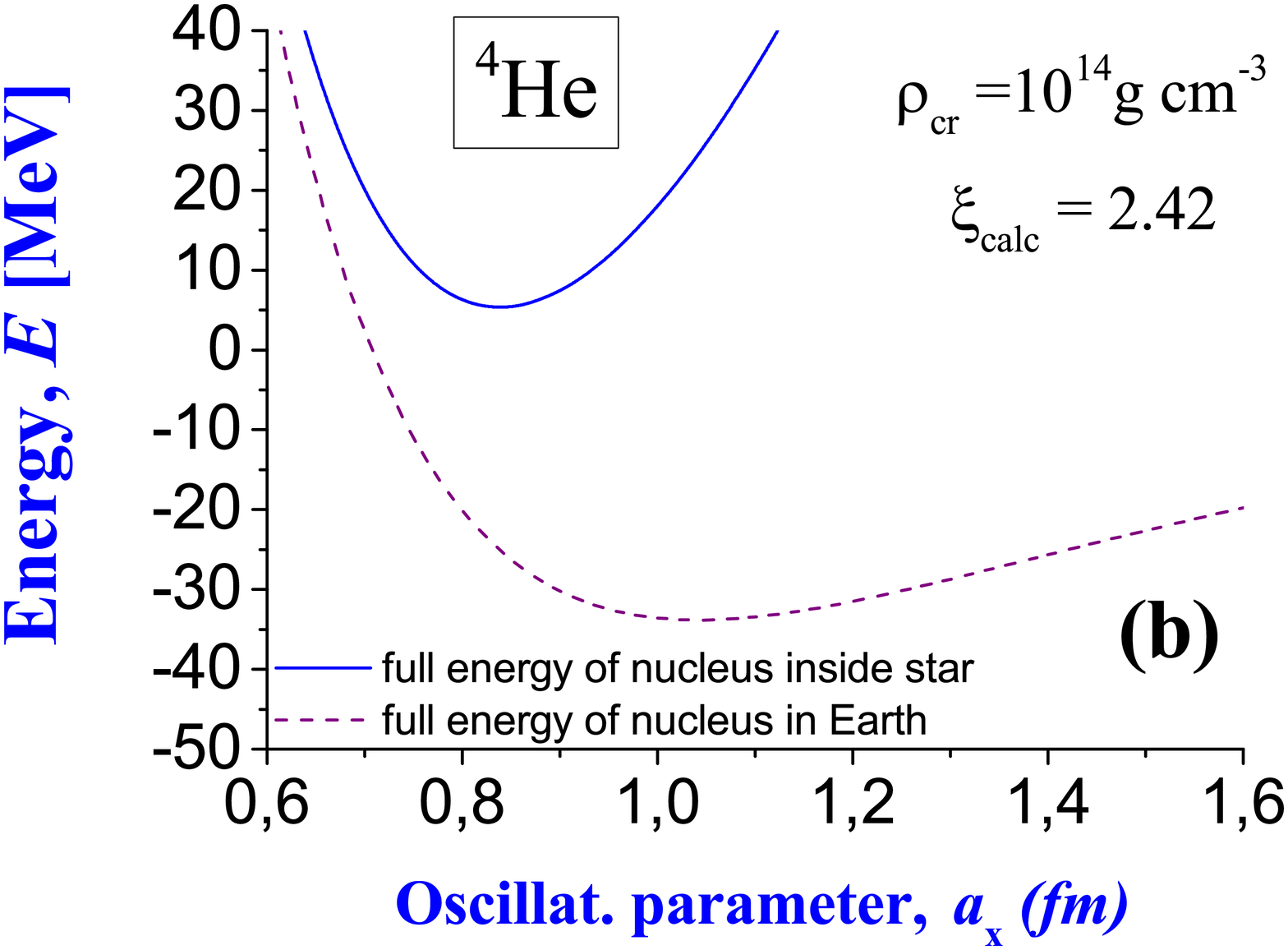}}
\vspace{-2mm}
\caption{\small (Color online)
Panel (a):
Energy of ``nucleus'' \isotope[4]{He} inside star at distance $\xi=2.42$ from center of star with density at center $\rho_{\rm cr} = 10^{14} {\rm g}\, {\rm cm}^{-3}$ (at $n=3$).
At minimum of the full energy of nucleus, we obtain:
$a = 0.8374$~fm,
$E_{\rm full} = 5.038$~MeV,
$E_{\rm full\, per\, nucl} = 1.327$~MeV,
$E_{\rm kin} = 138.096$~MeV,
$E_{\rm Coul} = 1.363$~MeV,
$E_{\rm nucl} = -164.492$~MeV,
$E_{\rm star} = 30.313$~MeV.
Panel (b): Full energy of nucleus inside star in comparison with full energy of this nucleus in Earth.
\label{fig.4}}
\end{figure}

\begin{figure}[htbp]
\centerline{\includegraphics[width=90mm]{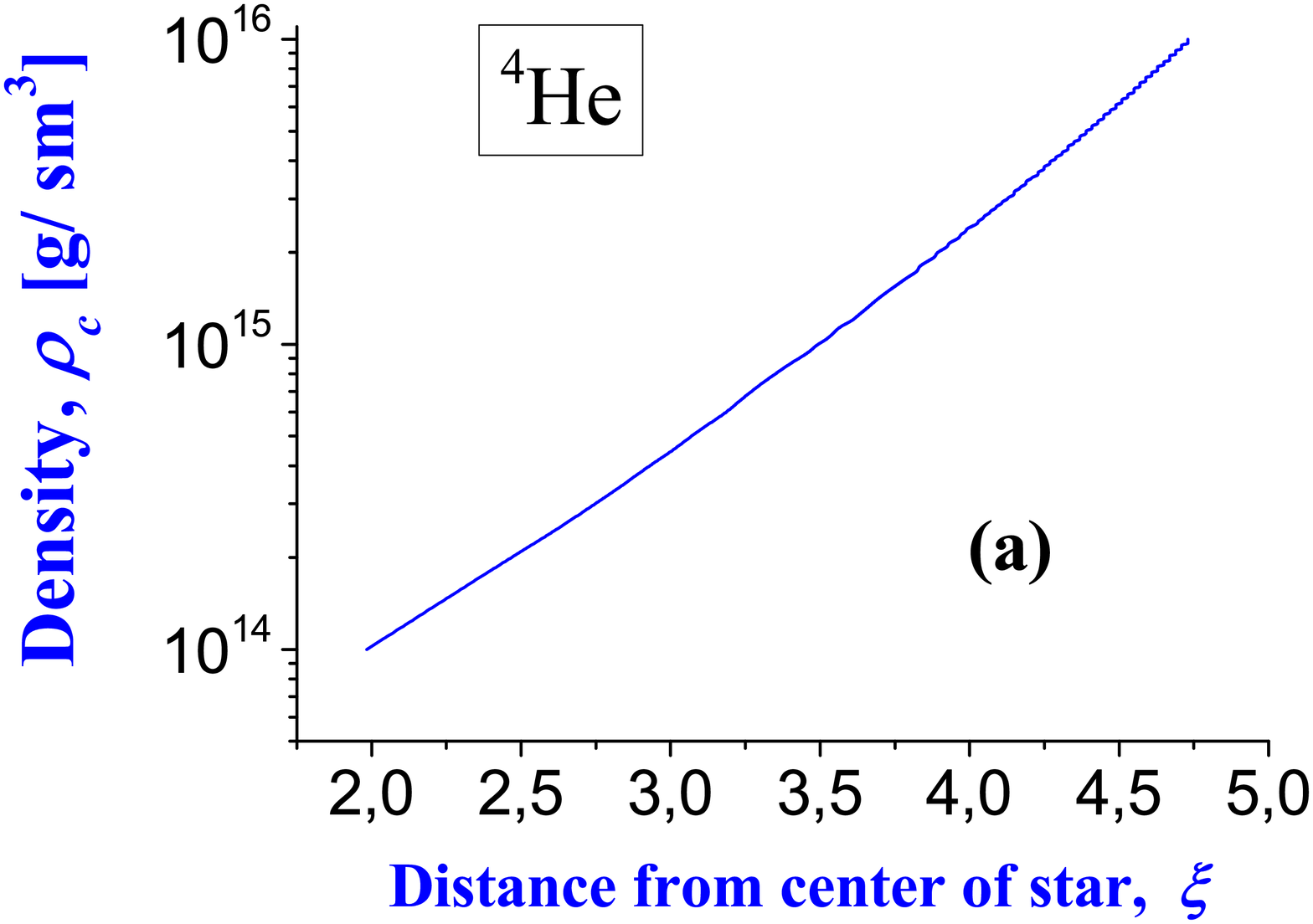}}
\vspace{-2mm}
\caption{\small (Color online)
Critical distance $\xi$ from center of star in dependence on its density at center, where disintegration of nucleus \isotope[4]{He} on nucleons takes place.
\label{fig.5}}
\end{figure}

Now we shall analyze possibility of nucleus to disintegrate on individual nucleons in the neutron star.
Different types of energy for nucleus \isotope[4]{He} concerning to its full energy are shown in Fig.~\ref{fig.4}.
Also one can see that full energy of nucleus is already positive. This means that system of nucleons representing the nucleus is not bound system (i.e. nucleus is disintegrated on nucleons).
%
%
Also one can see that parameter $a$, corresponding to the minimum of the full energy of system of nucleons, is decreased in comparison with its value for nucleus in usual conditions (i.e. outside from star).
So, relative distances between nucleons of \isotope[4]{He} is decreased. This is explained by influence of pressure of stellar medius on these nucleons.
The similar tendencies we obtain for other isotopes of \isotope{He} and \isotope{Be} considered above.
%
After analysis above, we would like to estimate where inside star there is such a phenomenon of disintegration of nucleus.
Such calculations are presented in Fig.~\ref{fig.5} in dependence on density of stellar medium at center of neutron star.
In particular, one can see that this model describes that for more compact stars dissosoiation of nuclei happens closer to external surface.

\section{Bremsstrahlung emission of photons during scattering of protons off nuclei in stellar medium of compact stars at $n=3$
\label{sec.brem.1}}

We analyze emission of photons in nuclear reactions inside compact stars.
We will focus on question, how a dense medium of star influences on emission of photons. Note that such a question has not been studied else.
Note that some calculations were done for proton-capture reactions~\cite{Maydanyuk_Zhang.2015.PRC}, that is enough popular for stars.
However, in those calculations nucleus was considered as stable, without influence of stellar medium.
Now we will take into account change of nucleus due to influence of stellar medium.
To be close to that analysis, we will choose scattering of protons of nuclei.

For scattering of proton off nucleus, we can rewrite hamiltonian (\ref{eq.star.2.1}) as
%
\begin{equation}
\begin{array}{lcl}
  \hat{H} =
  - \displaystyle\frac{\hbar^{2}}{2m} \displaystyle\sum\limits_{i=1}^{A+1} \grad_{i}^{2} +
  \displaystyle\sum\limits_{i,j=1}^{A+1} V_{\rm DOS} (|\vb{r}_{i} - \vb{r}_{j}|) +
  \displaystyle\sum\limits_{i,j=1}^{A+1} V_{\rm star} (|\vb{r}_{i} - \vb{r}_{j}|).
\end{array}
\label{eq.brem.1.1}
\end{equation}
Inclusion of emission of bremsstrahlung photons can be described via the following hamiltonian:
%
\begin{equation}
\begin{array}{lcl}
  \hat{H}_{full} =
  - \displaystyle\frac{\hbar^{2}}{2m} \displaystyle\sum\limits_{i=1}^{A+1} \grad_{i}^{2} +
  \displaystyle\sum\limits_{i,j=1}^{A+1} V_{\rm DOS} (|\vb{r}_{i} - \vb{r}_{j}|) +
  \displaystyle\sum\limits_{i,j=1}^{A+1} V_{\rm star} (|\vb{r}_{i} - \vb{r}_{j}|) +
  \hat{H}_{\gamma},
\end{array}
\label{eq.brem.1.2}
\end{equation}
where $\hat{H}_{\gamma}$ is a new operator which describes emission of bremsstrahlung photons for the studied reaction inside star.
A concrete form of this operator should be defined explicitly.
%

Emission of bremsstrahlung photons without influence of stellar medium [i.e. without last term in Eq.~(\ref{eq.brem.1.2})] for scattering of protons off nuclei in conditions of Earth was studied enough often by different researchers.
Here, agreement between theory and existed experimental information has been obtained with the highest precision for this reaction in frameworks of approach~\cite{Maydanyuk_Zhang.2015.PRC,Maydanyuk.2012.PRC}
(this is data~\cite{Clayton.1992.PRC} for $p+\isotope[208]{Pb}$ at proton energy beam of $E_{\rm p}=145$~MeV,
data~\cite{Goethem.2002.PRL} for $p+\isotope[12]{C}$, $p+\isotope[58]{Ni}$, $p+\isotope[107]{Ag}$, $p+\isotope[197]{Au}$ at proton energy beam of $E_{\rm p}=190$~MeV
and corresponding calculations in Figs.~5--8 in Refs.~\cite{Maydanyuk_Zhang.2015.PRC}).
Therefore, we will generalize bremsstrahlung formalism for these reactions from above-zero energies up to intermediate energies in stars, basing on formalism and results of papers~\cite{Maydanyuk_Zhang.2015.PRC,Maydanyuk.2012.PRC}
(see improvements of formalism in
Refs.~\cite{Maydanyuk_Zhang_Zou.2016.PRC,Maydanyuk_Zhang_Zou.2018.PRC,Maydanyuk_Zhang_Zou.2019.PRC.microscopy,Liu_Maydanyuk_Zhang_Liu.2019.PRC.hypernuclei,Maydanyuk_Zhang_Liu.2019.arxiv.alpha_nucleus},
for other reactions see Refs.~\cite{Maydanyuk_Zhang_Zou.2016.PRC,Maydanyuk.2009.JPS,Maydanyuk.2009.TONPPJ,Maydanyuk.2009.NPA,Maydanyuk.2010.PRC,Maydanyuk.2011.JPG,Maydanyuk.2011.JPCS,%
Maydanyuk.2006.EPJA,Maydanyuk.2008.EPJA,Maydanyuk.2008.MPLA}).


According to such an approach, for reaction in conditions of Earth in laboratory frame we define cross-section of bremsstrahlung emission of photons in frameworks of papers~\cite{Maydanyuk.2012.PRC},
%
%
where the full matrix element of emission of photons is defined as
\begin{equation}
  \langle \Psi_{f} |\, \hat{H}_{\gamma} |\, \Psi_{i} \rangle_{0} \;\; = \;\;
  \sqrt{\displaystyle\frac{2\pi\, c^{2}}{\hbar w_{\rm ph}}}\,
  \Bigl\{ M_{P} + M_{p}^{(E)} + M_{p}^{(M)} + M_{\Delta E} + M_{\Delta M} + M_{k} \Bigr\},
\label{eq.brem.1.3}
\end{equation}
matrix elements have form
%
\begin{equation}
\begin{array}{lll}
\vspace{0.2mm}
  M_{p}^{(E,\, {\rm dip},0)} & = &
  i \hbar^{2}\, (2\pi)^{3}
  \displaystyle\frac{e}{\mu c}\;
  Z_{\rm eff}^{\rm (dip, 0)}\;
  \displaystyle\sum\limits_{\alpha=1,2} \vb{e}^{(\alpha)} \cdot \vb{I}_{1}, \\

  M_{p}^{(M,\, {\rm dip},0)} & = &
  -\, \hbar\, (2\pi)^{3} \displaystyle\frac{1}{\mu}\;
  \vb{M}_{\rm eff}^{\rm (dip, 0)}
  \displaystyle\sum\limits_{\alpha=1,2} \Bigl[ \vb{I}_{1} \times \vb{e}^{(\alpha)} \Bigr], \\

  M_{\Delta E} & = & 0, \\

  M_{\Delta M} & = & i\, \hbar\, (2\pi)^{3}\: f_{1} \cdot |\vb{k}_{\rm ph}| \cdot z_{\rm A} \cdot I_{2}, \\

  M_{k} & = & \displaystyle\frac{f_{k}}{f_{1}} \cdot M_{\Delta M},
\end{array}
\label{eq.brem.1.4}
\end{equation}
coefficient are defined as
%
\begin{equation}
\begin{array}{lll}
  f_{1} = \displaystyle\frac{A-1}{2A}\: \mu_{\rm pn}^{\rm (an)}, &

  \displaystyle\frac{f_{k}}{f_{1}} =
  - \displaystyle\frac{\hbar A}{A-1}
\end{array}
\label{eq.brem.1.5}
\end{equation}
%
and integrals are defined as
%
\begin{equation}
\begin{array}{lll}
\vspace{0.5mm}
  \vb{I}_{1} = \biggl\langle\: \Phi_{\rm p - nucl, f} (\vb{r})\; \biggl|\, e^{-i\, \vb{k}_{\rm ph} \vb{r}}\; \vb{\displaystyle\frac{d}{dr}} \biggr|\: \Phi_{\rm p - nucl, i} (\vb{r})\: \biggr\rangle, \\
  I_{2} = \Bigl\langle \Phi_{\rm p - nucl, f} (\vb{r})\; \Bigl|\, e^{i\, c_{\rm p}\, \vb{k_{\rm ph}} \vb{r}}\, \Bigr|\, \Phi_{\rm p - nucl, i} (\vb{r})\: \Bigr\rangle.
\end{array}
\label{eq.brem.1.6}
\end{equation}
%
Here,
$\vb{r}$ is radius vector from center-of-mass on nucleus to the scattered proton,
$\mu = m_{\rm p}\, m_{A} / (m_{\rm p} + m_{A})$ is reduced mass,
$A$ is number of nucleons in nucleus,
$c_{\rm p} = m_{\rm p}/(m_{\rm p}+m_{A})$,
$\vb{e}^{(\alpha)}$ are unit vectors of polarization of the photon emitted [$\vb{e}^{(\alpha), *} = \vb{e}^{(\alpha)}$], $\vb{k}_{\rm ph}$ is wave vector of the photon and $w_{\rm ph} = k_{\rm ph} c = \bigl| \vb{k}_{\rm ph}\bigr|c$, $E_{\rm ph} = \hbar w_{\rm ph}$ is energy of photon.
Vectors $\vb{e}^{(\alpha)}$ are perpendicular to $\vb{k}_{\rm ph}$ in Coulomb calibration.
We have two independent polarizations $\vb{e}^{(1)}$ and $\vb{e}^{(2)}$ for the photon with impulse $\vb{k}_{\rm ph}$ ($\alpha=1,2$).
$\mu_{\rm pn}^{\rm (an)} = \mu_{\rm p}^{\rm (an)} + \mu_{\rm n}^{\rm (an)}$,
$\mu_{\rm p}^{\rm (an)}$ and $\mu_{\rm n}^{\rm (an)}$ are anomalous magnetic moments of proton and neutron.

%
The matrix elements $M_{p}^{(E,\, {\rm dip},0)}$ and $M_{p}^{(M,\, {\rm dip},0)}$ describe coherent bremsstrahlung emission of photons of electric and magnetic types,
the matrix elements $M_{\Delta E}$ and $M_{\Delta M}$ describe incoherent bremsstrahlung emission of photons of electric and magnetic types.
$M_{P}$ is related with motion of full nuclear system, which we neglect in this paper.
Effective electric charge and magnetic moment of system in dipole approximation (i.e. at $\vb{k_{\rm ph}} \vb{r} \to 0$) are
%
%
\begin{equation}
\begin{array}{lll}
  Z_{\rm eff}^{\rm (dip, 0)} = \displaystyle\frac{m_{A}\, z_{\rm p} - m_{\rm p}\, z_{\rm A}}{m_{\rm p} + m_{A}}, &
  \textbf{M}_{\rm eff}^{\rm (dip,0)} = - \displaystyle\frac{m_{\rm p}}{m_{\rm p} + m_{A}}\, \vb{M}_{A}.
\end{array}
\label{eq.brem.1.7}
\end{equation}
$m_{\rm p}$ and $z_{\rm p}$ are mass and charge of proton,
$m_{A}$ and $z_{A}$ are mass and charge of nucleus.
Here, we introduced magnetic moment of nucleus $\vb{M}_{A}$ (without inclusion of characteristics of photons emitted):
%
%
\begin{equation}
\begin{array}{lll}
  \vb{M}_{A} =
  \displaystyle\sum_{j=1}^{A}
    \Bigl\langle \psi_{\rm nucl, f} (\beta_{A})\, \Bigl|\, \mu_{j}^{\rm (an)}\, m_{Aj}\, \sigmabf \Bigr| \psi_{\rm nucl, i} (\beta_{A}) \Bigr\rangle,
\end{array}
\label{eq.brem.1.8}
\end{equation}
where $\mu_{j}^{\rm (an)}$ is anomalous magnetic moment of proton or neutron in nucleus,
$m_{Aj}$ is mass of nucleon with number $j$ in nucleus,
$\sigmabf$ is operator of spin (acting on wave function of nucleon of nucleus).

For first estimations of bremsstrahlung emission for nuclear reactionbs in stellar medium, we shall us perturbation theory.
We will take into account influence of stellar medium on emission as
%
%
\begin{equation}
\begin{array}{lll}
  \hat{H}_{\gamma\, new} = \hat{H}_{\gamma 0} + \Delta \hat{H}_{\gamma}, &
  \Delta \hat{H}_{\gamma} = \displaystyle\sum\limits_{i,j=1}^{A+1} V_{\rm star} (|\vb{r}_{i} - \vb{r}_{j}|),
\end{array}
\label{eq.brem.1.9}
\end{equation}
From here we find the matrix element of emission for reaction inside star as
%
\begin{equation}
  \langle \Psi_{f} |\, \hat{H}_{\gamma} |\, \Psi_{i} \rangle_{\rm star} \;\; = \;\;
  \langle \Psi_{f} |\, \hat{H}_{\gamma} |\, \Psi_{i} \rangle_{0} + \langle \Psi_{f} |\, \Delta \hat{H}_{\gamma} |\, \Psi_{i} \rangle.
\label{eq.brem.1.10}
\end{equation}
According to perturbation theory,
for determination of the first correction we should use wave functions of unperturbed system, i.e. we take wave functions as in matrix element~(\ref{eq.brem.1.3}):
%
\begin{equation}
\begin{array}{lcl}
  \langle \Psi_{f} |\, \Delta \hat{H}_{\gamma} |\, \Psi_{i} \rangle =
  \sqrt{\displaystyle\frac{2\pi\, c^{2}}{\hbar w_{\rm ph}}}\, \cdot M_{\rm star} (E_{\rm ph}), &

  M_{\rm star} (E_{\rm ph}) =
  N \cdot F_{P} (R) \cdot
  \displaystyle\int \varphi_{\rm p-nucl}^{2} (\vb{r}, k_{f})\, \varphi_{0}^{2} (\vb{r})\, |\vb{r}|\; \vb{dr},
\end{array}
\label{eq.brem.1.11}
\end{equation}
where
%
\begin{equation}
\begin{array}{lcl}
  F_{P} (R) = P (R) = K \cdot \rho^{\gamma} (R).
\end{array}
\label{eq.brem.1.12}
\end{equation}
Here $\varphi_{\rm p-nucl} (\vb{r}, k_{f})$ is wave function of scattering of proton off nucleus (energy has continuous spectrum, as emitted photon takes some energy of proton-nucleus system),
$\varphi_{0}^{2} (\vb{r})$ is wave function of nucleus (energy has only discrete levels, nucleons are only in bound states).
We additionally renormalize wave function of scattering of proton off nucleus%
\footnote{This is caused by that normalization of wave function of photon is determined by factor $\sqrt{\displaystyle\frac{2\pi\, c^{2}}{\hbar w_{\rm ph}}}$
at exponent of vector potential of electromagnetic field $\vb{A}$ in QED
[see representation~(5) in Ref.~\cite{Maydanyuk.2012.PRC}], that in principle is different from normalization of wave functions of nucleons in bound states and states of scattering in quantum mechanics.}.
%
%
For \isotope[4]{He} we have
%
\begin{equation}
\begin{array}{llll}
  \varphi_{0} (\vb{r}) = \varphi_{\rm n_{x}=0} (x) \cdot \varphi_{\rm n_{y}=0} (y) \cdot \varphi_{\rm n_{z}=0} (z), &
  \varphi_{n_{x}=0} (x) =
  \displaystyle\frac{\exp{- \displaystyle\frac{x^{2}}{2\,a^{2}}}}{\pi^{1/4}\, \sqrt{2^{n_{x}\, n_{x}!}}\, \sqrt{a}} \cdot H_{\rm n_{x}=0} \Bigl(\displaystyle\frac{x}{a}\Bigr) =
  \displaystyle\frac{\exp{- \displaystyle\frac{x^{2}}{2\,a^{2}}}}{\pi^{1/4}\, \sqrt{a}}, &
  N=12.
\end{array}
\label{eq.brem.1.13}
\end{equation}
Substituting Eq.~(\ref{eq.brem.1.13}) to (\ref{eq.brem.1.11}) for \isotope[4]{He} (at $a=b=c$), we obtain:
%
\begin{equation}
\begin{array}{lcl}
  M_{\rm star} (E_{\rm ph}) =
  F_{P} (R) \cdot \displaystyle\frac{N}{\pi^{3/2}\, a^{3}}
  \displaystyle\int \varphi_{\rm p-nucl}^{2} (\vb{r}, k_{f})\,
  \exp{- \displaystyle\frac{r^{2}}{a^{2}}}\, r\; \vb{dr},
\end{array}
\label{eq.brem.1.14}
\end{equation}

We calculate the wave functions $\varphi_{\rm p-nucl}$ numerically concerning the chosen potential
of interaction between the proton and the spherically symmetric core.
For description of proton-nucleus interaction we use the potential as
$V (r) = v_{c} (r) + v_{N} (r) + v_{\rm so} (r) + v_{l} (r)$,
where $v_{c} (r)$, $v_{N} (r)$, $v_{\rm so} (r)$ and $v_{l} (r)$ are Coulomb,
nuclear, spin-orbital and centrifugal components
defined in Ref.~\cite{Becchetti.1969.PR}.

\begin{figure}[htbp]
\centerline{\includegraphics[width=90mm]{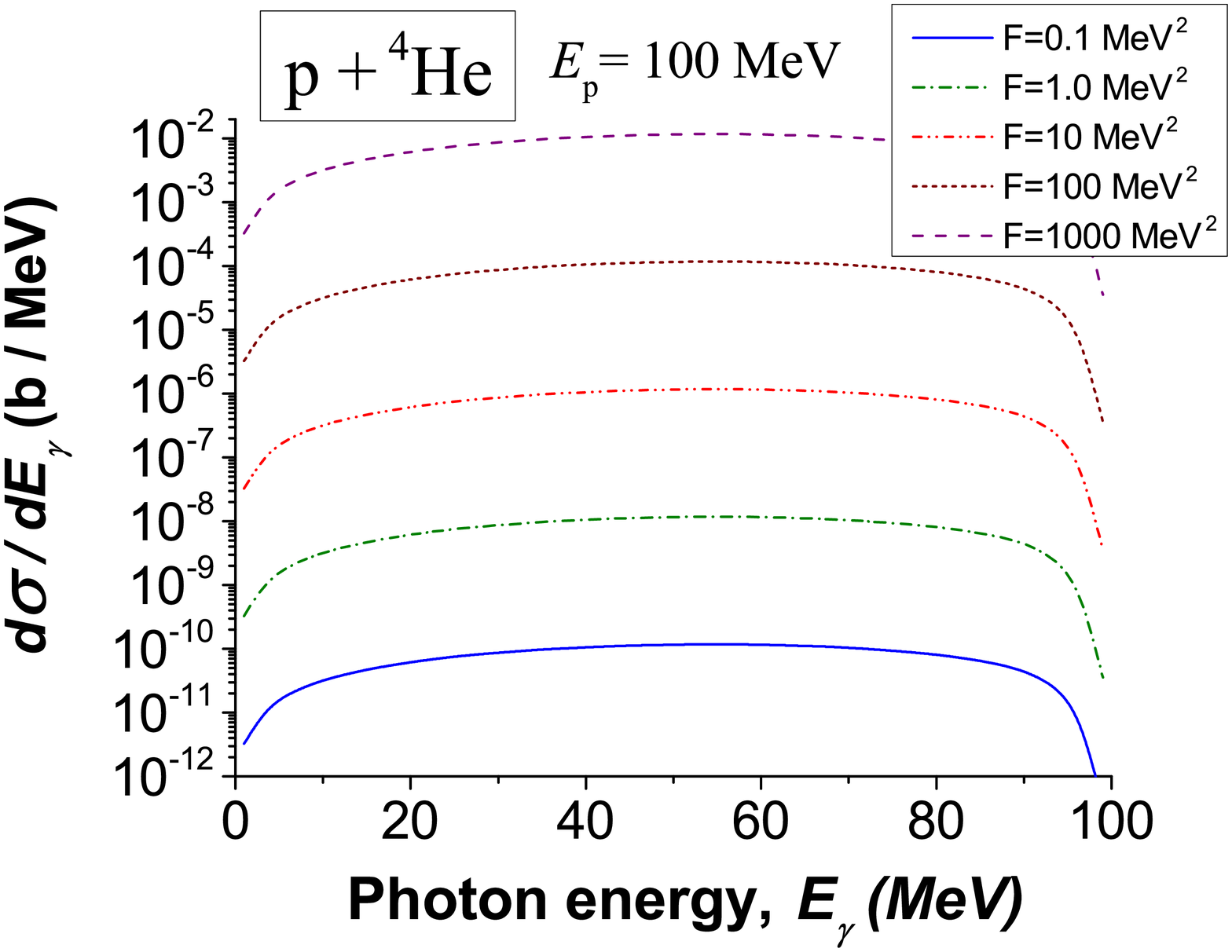}
\hspace{-1mm}\includegraphics[width=88mm]{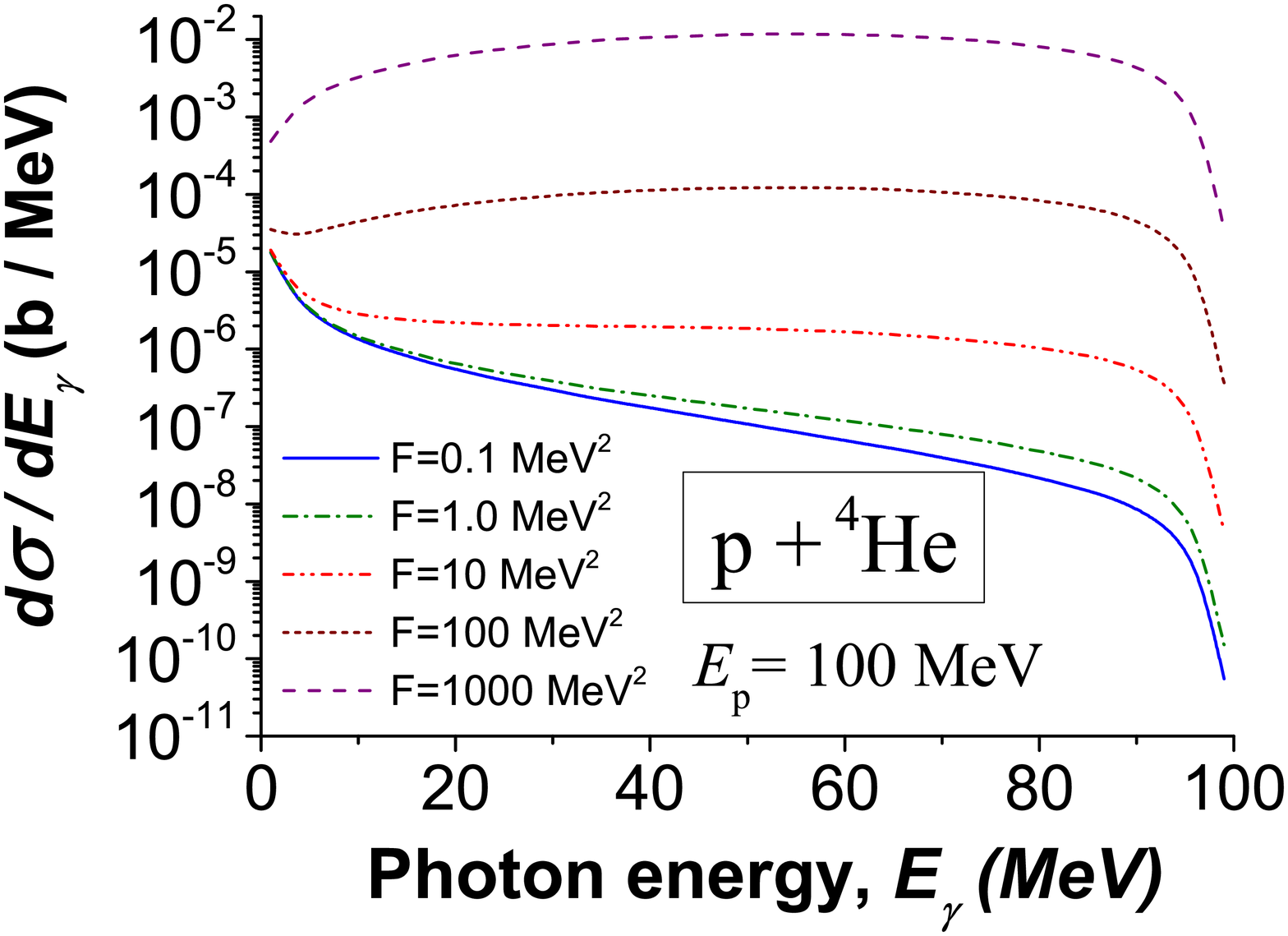}}
\vspace{-3mm}
\caption{\small (Color online)
Bremsstrahlung emission of photons in scattering of protons off nuclei \isotope[4]{He} inside star at energy of protons of $E_{\rm p}=100$~MeV
[
we calculate spectrum on the basis of the leading matrix element $M_{p}^{(E,\, {\rm dip},0)}$,
which gives the largest contribution to full spectrum, according to analysis in Refs.~\cite{Maydanyuk_Zhang.2015.PRC,Maydanyuk.2012.PRC}
%
].
Contribution on the basis of matrix element
$\langle \Psi_{f} |\, \Delta \hat{H}_{\gamma} |\, \Psi_{i} \rangle$ in Eq.~(\ref{eq.brem.1.10}) (a),
and full spectrum on the basis of matrix element $\langle \Psi_{f} |\, \hat{H}_{\gamma} |\, \Psi_{i} \rangle_{\rm star}$ in Eq.~(\ref{eq.brem.1.10}) (b) are shown in these figures.
\label{fig.9}}
\end{figure}

Results of calculation of spectrum of emision of bremsstrahlung photons in scattering of protons off nuclei in stars on the basis of such an approach are shown in Fig.~\ref{fig.9}.
%
From this figure we conclude the following.
\begin{itemize}
\item
In the white dwarfs, according to Fig.~\ref{fig.4}, influence of stellar medium on emission is not larger than $0.1~MeV^{2}$.
This means that influence of stellar medium imperceptibly affects on emission of bremsstrahlung photons.
In particular, such a conclusion can be formulated for nuclear reactions inside Sun.
I.e., it turns out that we have enough accurate description of emission of bremsstrahlung photons during nuclear reactions in Sun, white dwarfs and similar stars.


\item
For neutron stars, influence of stellar medium is essentially more intensive and it crucially changes shape of the spectrum of the bremsstrahlung photons (see Fig.~\ref{fig.9}).
In the simplest approximation, one can find that maximum of probability of the emitted photons is for their energy,
which is half of energy of the scattered protons: $E_{\rm ph} \simeq E_{\rm p} / 2$.
One can see that the most intensive emission is created in the bowel of the star, while the weakest emission is from the periphery (for the same energy of the scattered proton).

\end{itemize}

\begin{figure}[htbp]
\centerline{\includegraphics[width=90mm]{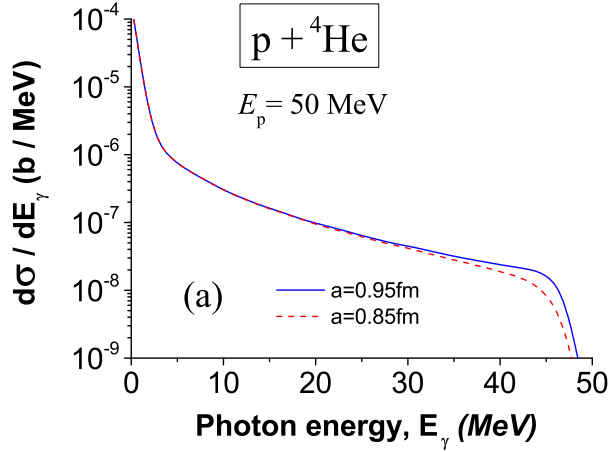}
}
\vspace{-2mm}
\caption{\small (Color online)
Cross-sections of bremsstrahlung photons of electric type emitted during the scattering of protons off \isotope[4]{He} inside star (red dashed line) and in vacuum (blue solid line)
[parameters of calculations: $r_{\rm max} = 20000$~fm, 10000 intervals of integration of matrix elements of emission
].
One can see small difference between the spectra at high energy part of photons.
\label{fig.10}}
\end{figure}

In Fig.~\ref{fig.10} one can see spectra of the bremsstrahlung photons emitted in $p + \isotope[4]{He}$ is star in comparison with the same reaction in vacuum.
In such calculations, we just use deformation of nucleus due to influence of stellar medium (for convenience we chose stage before disintegration of nucleus, and obtain parameter $a=0.85$~fm, in vacuum we have $a=0.95$~fm)
without inclusion of contribution $\langle \Psi_{f} |\, \Delta \hat{H}_{\gamma} |\, \Psi_{i} \rangle$.
One can see that such an influence is almost neglectable in comparison with influence of term $\langle \Psi_{f} |\, \Delta \hat{H}_{\gamma} |\, \Psi_{i} \rangle$ given in Fig.~\ref{fig.9}.



\section{Conclusions
\label{sec.conclusions}}

In this paper we investigate ability (role) of nuclear forces to combine nucleons as bound nuclear system in dependence on its deep location inside the compact star.
In order to perform such a research, we generalize the model of deformed oscillator shells \cite{Steshenko.1971.YF,Steshenko.1970.UPJ,Steshenko.1976.preprint} with two nucleon forces with new inclusion of additional influence of stellar medium.
We have obtained new simple exact formulas of energy for the lightest even-even nuclei,
that is convenient for analysis of stellar influence on binding energy of nuclei.
As studied star, polytropic stars at $n=3$ with densities characterized from white dwarf to neutron star were included to analysis.

We observe a phenomenon of dissociation of nucleus --- its disintegration on individual nucleons, starting from some critical distance between this nucleus and center of star with high density.
We explain this phenomenon by the following logic.
Forces of stellar medium press on nucleons of nucleus.
The deeper this nucleus is located in star, the stronger such forces press on nucleus.
However, binding energy (it is negative for nucleus in the external layer of star) is increased at deeper location of this nucleus in star.
Starting from some critical distance from nucleus to center of star, the binding energy becomes positive (see Fig.~\ref{fig.4}).
This means that full energy of individual nucleons of the studied nucleus is already larger than mass of this nucleus,
i.e. we obtain unbound system of nucleons and nucleus is disintegrated on nucleons.
According to estimations, we observe such a phenomenon in neutron stars, while in white dwarfs its is not observed.
We have estimated such a critical distance for nucleus \isotope[4]{He} in dependence on density at center of neutron star (see Fig.~\ref{fig.5}), where disintegration of this nucleus on nucleons takes place.
%
The kinetic energy is increased at deeper location of nucleus in star.
At decreasing distance from the studied nucleus to center of star, change of kinetic energy is unlimited, while change of nuclear energy is limited.
So, ratio between kinetic energy of nucleons of nucleus and nuclear energy of nucleus is changed also.
%

Basing on the model above, we have generalized the bremsstrahlung formalism \cite{Maydanyuk_Zhang.2015.PRC,Maydanyuk.2012.PRC} (see also improvements of this formalism in
Refs.~\cite{Maydanyuk_Zhang_Zou.2016.PRC,Maydanyuk_Zhang_Zou.2018.PRC,Maydanyuk_Zhang_Zou.2019.PRC.microscopy,Liu_Maydanyuk_Zhang_Liu.2019.PRC.hypernuclei,Maydanyuk_Zhang_Liu.2019.arxiv.alpha_nucleus})
for scattering of protons off nuclei in compact stars.
Using such a new model, we find the following.
(1) In the white dwarfs, influence of stellar medium imperceptibly affects on emission of bremsstrahlung photons.
This means that we have enough accurate description of emission of bremsstrahlung photons during nuclear reactions in Sun, white dwarfs and similar stars.
%
%
(2) For neutron stars, influence of stellar medium is essentially more intensive and it crucially changes shape of the spectrum of the bremsstrahlung photons.
Maximum of probability of the emitted photons is for their energy, which is half of energy of the scattered protons: $E_{\rm ph} \simeq E_{\rm p} / 2$.
The most intensive emission is created in the bowel of the star, while the weakest emission is from the periphery (for the same energy of the scattered proton).

Summarizing, we find the model of deformed oscillator shells as convenient and not complicated technically basis for obtaining clear understanding about different forces and emission of bremsstrahlung photons in nuclear reactions in compact stars.

\section*{Acknowledgements
\label{sec.acknowledgements}}

S.~P.~M. is highly appreciated to Dr. A.~I.~Steshenko for deep insight to the DOS model and help.
Authors are highly appreciated to Profs.~V.~S.~Vasilevsky, M.~I.~Gorenstein, A.~V.~Nesterov for useful recommendations and interesting discussions concerning to modern many-nucleons nuclear models and
physics of nuclear processes inside dense stellar medium.
Authors also highly appreciated to 
Prof.~Janos~Balog for interesting discussions concerning to physics of nucleon-nucleon interactions, and
Prof.~Zhigang Xiao for interesting discussions concerning to emission of bremsstrahlung in heavy-ion collisions.

\appendix
\section{Correction of energy of nucleus due to influence of stellar medium
\label{sec.app.1}}


In this Section we shall find correction of energy of nucleus due to influence of stellar medium (\ref{eq.star.2.3}):
\begin{equation}
\begin{array}{lcl}
  \Delta E_{\rm star} & = &
  \Bigl\langle \Psi(1 \ldots A)
    \Bigl|
      \displaystyle\sum\limits_{i,j=1}^{A} V_{\rm star} (R, \vb{r}_{i}, \vb{r}_{j})
    \Bigl| \Psi(1 \ldots A) \Bigr\rangle.
\end{array}
\label{eq.app.1.1}
\end{equation}
Substituting Eq.~(\ref{eq.star.2.2}) to this formula and taking into account the same action of force $\vb{F}_{P}(R)$ for each nucleon, we obtain:
\begin{equation}
\begin{array}{lcl}
  \Delta E_{\rm star} & = &

  \vb{F}_{P} (R) \cdot
  \displaystyle\sum\limits_{i,j=1}^{A}
  \Bigl\langle \Psi(1 \ldots A)
    \Bigl|
      \vb{r}_{i} - \vb{r}_{j}
    \Bigl| \Psi(1 \ldots A) \Bigr\rangle.
\end{array}
\label{eq.app.1.2}
\end{equation}
We use property:
%
%
\begin{equation}
\begin{array}{lcl}
\vspace{1mm}
  \langle \Psi_{f} (1 \cdots A )\, |\, \hat{V}\, (\vb{r}_{i}, \vb{r}_{j}) |\, \Psi_{i} (1 \cdots A ) \rangle =  \\
 = \quad
  \displaystyle\frac{1}{A\,(A-1)}\;
  \displaystyle\sum\limits_{k=1}^{A}
  \displaystyle\sum\limits_{m=1, m \ne k}^{A}
  \biggl\{
    \langle \psi_{k}(i)\, \psi_{m}(j) |\, \hat{V}\, (\vb{r}_{i}, \vb{r}_{j}) |\, \psi_{k}(i)\, \psi_{m}(j) \rangle -
    \langle \psi_{k}(i)\, \psi_{m}(j) |\, \hat{V}\, (\vb{r}_{i}, \vb{r}_{j}) |\, \psi_{m}(i)\, \psi_{k}(j) \rangle
  \biggr\}.
\end{array}
\label{eq.app.1.3}
\end{equation}
Here, summation is performed over all states for the given configuration of nucleus (they are denoted by indexes $m$ and $k$).
All nuclerons are numeberd by indexes $i$ and $j$.
We use representation for one-nucleon wave function:
%
\begin{equation}
  \psi_{\lambda_{s}} (s) =
  \varphi_{n_{s}} (\mathbf{r}_{s})\,
  \bigl|\, \sigma^{(s)} \tau^{(s)} \bigr\rangle,
\label{eq.app.1.4}
\end{equation}
where
$\varphi_{n_{s}}$ is space function of the nucleon with number $s$,
$n_{s}$ is number of state of the space function of the nucleon with number $s$,
$\bigl|\, \sigma^{(s)} \tau^{(s)} \bigr\rangle$ is spin-isospin function of the nucleon with number $s$.
For operator $\hat{V}\, (\vb{r}_{i}, \vb{r}_{j})$ acting on space functions for two nucleons only, we calculate matrix element:
\begin{equation}
\begin{array}{lcl}
\vspace{0.5mm}
  \langle \Psi_{f} (1 \cdots A )\, |\, \hat{V}\, (\vb{r}_{i}, \vb{r}_{j}) |\, \Psi_{i} (1 \cdots A ) \rangle =

  \displaystyle\frac{1}{A\,(A-1)}\;
  \displaystyle\sum\limits_{k=1}^{A}
  \displaystyle\sum\limits_{m=1, m \ne k}^{A}\;
  \biggl\{
    \Bigl\langle
      \varphi_{k} (\vb{r}_{i})\, \varphi_{m} (\vb{r}_{j}) \Bigl|\, \hat{V}\, (\vb{r}_{i}, \vb{r}_{j}) \Bigr|\, \varphi_{k} (\vb{r}_{i})\, \varphi_{m} (\vb{r}_{j})
    \Bigr\rangle\; - \\
\vspace{3mm}
  - \quad
  \Bigl\langle
    \varphi_{k} (\vb{r}_{i})\, \varphi_{m} (\vb{r}_{j}) \Bigl|\, \hat{V}\, (\vb{r}_{i}, \vb{r}_{j}) \Bigr|\, \varphi_{m} (\vb{r}_{i})\, \varphi_{k} (\vb{r}_{j})
  \Bigr\rangle\:
    \bigl\langle \sigma^{(k)} (i) \bigl|\, \sigma^{(m)} (i) \bigr\rangle\, \bigl\langle \sigma^{(m)} (j) \bigl|\, \sigma^{(k)} (j) \bigr\rangle\:
    \bigl\langle \tau^{(k)} (i) \bigl|\, \tau^{(m)} (i) \bigr\rangle\, \bigl\langle \tau^{(m)} (j) \bigl|\, \tau^{(k)} (j) \bigr\rangle \biggr\},
\end{array}
\label{eq.app.1.5}
\end{equation}
where orthogonalization properties of spin and isospin functions are used:
\begin{equation}
\begin{array}{llll}
  \bigl\langle \sigma^{(k)} (i) \bigl|\, \sigma^{(k)} (i) \bigr\rangle = 1, &
  \bigl\langle \tau^{(k)} (i) \bigl|\, \tau^{(k)} (i) \bigr\rangle = 1.
\end{array}
\label{eq.app.1.6}
\end{equation}


In particular, for \isotope[4]{He} Eqs.~(\ref{eq.app.1.6}) are simplified:
\begin{equation}
\begin{array}{llll}
  \bigl\langle \sigma^{(k)} (i) \bigl|\, \sigma^{(m)} (i) \bigr\rangle = \delta_{km}, &
  \bigl\langle \tau^{(k)} (i) \bigl|\, \tau^{(m)} (i) \bigr\rangle = \delta_{km},
\end{array}
\label{eq.app.1.7}
\end{equation}
and we obtain
\begin{equation}
\begin{array}{lcl}
  \langle \Psi (\isotope[4]{He})\, |\, \vb{r}_{i} - \vb{r}_{j} |\, \Psi (\isotope[4]{He}) \rangle =
  \displaystyle\int F_{0}^{2} (\vb{r}_{i}, \vb{r}_{j})\, (\vb{r}_{i} - \vb{r}_{j})\; \vb{dr}_{1}\, \vb{dr}_{2},
\end{array}
\label{eq.app.1.8}
\end{equation}
where Eq.~(\ref{eq.calc.densities.1.2}) for $F_{0} (\vb{r}_{i}, \vb{r}_{j})$ is used.
For correction of energy, from (\ref{eq.app.1.1}) we obtain:
\begin{equation}
\begin{array}{lcl}
  \Delta E_{\rm star} (\isotope[4]{He}) & = &
  \vb{F}_{P} (R) \cdot
    \displaystyle\sum\limits_{i,j=1}^{A=4}
    \Bigl\langle \Psi(\isotope[4]{He}) \Bigl| \vb{r}_{i} - \vb{r}_{j} \Bigl| \Psi(\isotope[4]{He}) \Bigr\rangle =


  12 \cdot \vb{F}_{P} (R) \cdot \displaystyle\int F_{0}^{2} (\vb{r}_{1}, \vb{r}_{2})\, (\vb{r}_{1} - \vb{r}_{2})\; \vb{dr}_{1}\, \vb{dr}_{2}.
\end{array}
\label{eq.app.1.9}
\end{equation}

In the spherically symmetric case ($a=b=c$), we calculate integral:
\begin{equation}
  \displaystyle\int F_{0}^{2} (\vb{r}_{1}, \vb{r}_{2})\, |\vb{r}_{1} - \vb{r}_{2}|\; \vb{dr}_{1}\, \vb{dr}_{2} =
  \displaystyle\frac{1}{\pi^{3}\, a^{6}}\,
  \displaystyle\int
    \exp\Bigl[ - \displaystyle\frac{x_{1}^{2} + x_{2}^{2} + y_{1}^{2} + y_{2}^{2} + z_{1}^{2} + z_{2}^{2}}{a^{2}} \Bigr]
    \cdot r_{12}\; \vb{dr}_{1}\, \vb{dr}_{2} =
  \displaystyle\frac{2^{3/2}\, a}{\pi^{1/2}}
\label{eq.app.1.10}
\end{equation}
and obtain solution:
\begin{equation}
\begin{array}{lcl}
  \Delta E_{\rm star} (\isotope[4]{He}) & = &
  12 \cdot F_{P} (R) \cdot \displaystyle\int F_{0}^{2} (\vb{r}_{1}, \vb{r}_{2})\, \abs{\vb{r}_{1} - \vb{r}_{2}}\; \vb{dr}_{1}\, \vb{dr}_{2} =
  \displaystyle\frac{12 \cdot 2^{3/2}\, a}{\pi^{1/2}} \cdot F_{P} (R).
\end{array}
\label{eq.app.1.11}
\end{equation}


\end{document}